\documentclass[twocolumn,superscriptaddress,letter]{revtex4}%
\usepackage{amssymb}
\usepackage{color}
\usepackage{graphicx}
\usepackage{dcolumn}
\usepackage{bm}
\usepackage[header,title,page,titletoc]{appendix}
\usepackage{amsmath}
\usepackage{amsfonts}
\usepackage{epsfig}
\usepackage[dvipdfm,pdfstartview=FitH,CJKbookmarks=true,
bookmarksnumbered=true, bookmarksopen=true,linktocpage=true,
colorlinks=true,pdfborder=001,citecolor=blue,
urlcolor=blue,linkcolor=blue,anchorcolor=blue, ]{hyperref}%
\providecommand{\U}[1]{\protect \rule{.1in}{.1in}}

\begin{document}
\title{Physics of $\mathcal{PT}$-Symmetric Quantum Systems at Finite Temperature}
\author{Qian Du}
\thanks{Who has same contribution to this work}
\affiliation{Center for Advanced Quantum Studies, Department of Physics, Beijing Normal
University, Beijing 100875, China}
\author{Kui Cao}
\thanks{Who has same contribution to this work}
\affiliation{Center for Advanced Quantum Studies, Department of Physics, Beijing Normal
University, Beijing 100875, China}
\author{Su-Peng Kou}
\thanks{Corresponding author}
\email{spkou@bnu.edu.cn}
\affiliation{Center for Advanced Quantum Studies, Department of Physics, Beijing Normal
University, Beijing 100875, China}

\begin{abstract}
We study parity-time-symmetric non-Hermitian quantum systems at finite
temperature, where the Boltzmann distribution law fails to hold. To
characterize their abnormal physical properties, a new quantum statistics
theory (the so-called quantum Liouvillian statistics theory) was developed, in
which the Boltzmann distribution law was replaced by the Liouvillian-Boltzmann
distribution law. Using it, we derived analytical results of thermodynamic
properties for thermal $\mathcal{PT}$ systems and found that a ``continuous''
thermodynamic phase transition occurs at the exceptional point, where a
zero-temperature anomaly exists.

\end{abstract}

\pacs{11.30.Er, 75.10.Jm, 64.70.Tg, 03.65.-W}
\maketitle

In statistical mechanics, the \emph{Boltzmann distribution (BZ) law} plays a
central role and governs the equilibrium distribution of different equilibrium
states at a particular temperature. According to it, a system will be in a
certain state as a function of its energy $E_{n}$ and of its temperature $T$.
As a result, the weights of different (quantum) states obey the BZ law, i.e. 
$P_{n}^{T}\sim e^{-E_{n}/k_{B}T}$. This universal distribution law was derived
by Boltzmann through an axiomatic way, which involves finding the most likely macrostates for a given the total energy under the assumption that all possible microstates were equally likely to occur. With the help of the BZ law, one can
recognize properties of macroscopic quantities of different physical systems
at finite temperature (finite-T).

The non-Hermitian (NH) problem in controlled open quantum systems has recently
begun to be considered one of the frontiers of physics. A parity-time
($\mathcal{PT}$)-symmetric NH quantum model was proposed by Bender and
Boettcher \cite{Bender98,Bender02,Bender07} in 1998, where $\mathcal{PT}%
$-symmetry spontaneous breaking ($\mathcal{PT}$-SSB) occurs at a critical
point (the so-called \textquotedblleft exceptional point \textquotedblright%
(EP)). $\mathcal{PT}$-symmetry systems have attracted a lot of researches in
different fields \cite{Giorgi10,Ge13,Hong2013,Korff07,Korff08} and various
approaches were proposed for realizing $\mathcal{PT}$-symmetric NH
models\cite{R10,Hang13,Guo09,Peng16,Zhang16,Chong11,Alois12,Liang13,Martin15,S16,Hossein17,S19,Feng13,luo,Bender13b,Joseph11,Assawaworrarit17,Choi18,Bittner12,Bo14,Poli15,Liu16,Feng14,Peng14,Hodaei14,
Zhu14,Popa14,Fleury15,Kun16,Wu19}. However, people are still in the dark about
the properties of a $\mathcal{PT}$-symmetric NH quantum system at finite-T
(the so-called \emph{thermal }$\mathcal{PT}$\emph{ systems}) and nothing is
known about their thermodynamic behavior. Hence, immediate questions appear,
such as: \emph{Do the thermal }$\mathcal{PT}$\emph{ systems still obey the BZ
law? If not, what distribution do they obey? Are there new physical phenomena
compared with their Hermitian counterparts for thermal }$\mathcal{PT}$
\emph{systems?}

In this letter, by taking a designed thermal $\mathcal{PT}$ system as an
example, we studied this issue and derived reliable results by solving the
quantum master equation. According to them, a surprising discovery is that the
BZ law is no longer applicable in such systems. As a result, a new theory
beyond the usual quantum statistical one is developed in order to completely
understand these abnormal physical properties of thermal $\mathcal{PT}$ systems.

\begin{figure}[ptb]
\includegraphics[clip,width=0.35\textwidth]{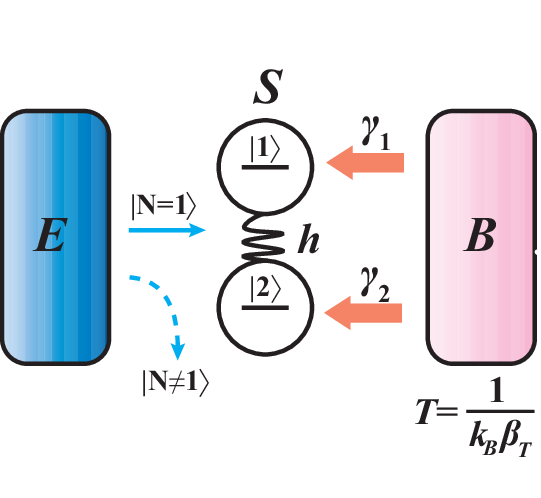}\caption{ Illustration of
a thermal $\mathcal{PT}$ system that comes from the controlled open quantum
system \textrm{S} coupling to two separated environments \textrm{B} and
\textrm{E}. \textrm{S} denotes a tight-binding model of two lattice sites, $1$
and $2$, with single fermionic particle; \textrm{B} denotes a thermal
environment with temperature $T=\frac{1}{k_{B}\beta_{T}}$; and \textrm{E}
denotes an environment under postselection with fixed particle number of
\textrm{S} to be $1$.}%
\end{figure}

To design a thermal $\mathcal{PT}$ system we couple a thermal bath to a
$\mathcal{PT}$ system. To reach this goal, we consider a controlled open
quantum system \textrm{S} coupling to two separated environments \textrm{B}
and \textrm{E}. (Fig.1) \textrm{B} could thermalize \textrm{S} to its
temperature $T=\frac{1}{k_{B}\beta_{T}}$ (we call it \emph{thermal
environment}), and \textrm{E} could lead to a relevant NH term for \textrm{S
}under postselection with the fixed particle number of \textrm{S} equal to $1$
(we call it \emph{NH environment}). The Hamiltonian of the total system is
given by
\begin{align}
\hat{H}  &  =\hat{H}_{\mathrm{S}}\otimes \hat{I}_{\mathrm{E}}\otimes \hat
{I}_{\mathrm{B}}+\hat{I}_{\mathrm{S}}\otimes \hat{H}_{\mathrm{E}}\otimes \hat
{I}_{\mathrm{B}}\nonumber \\
&  +\hat{I}_{\mathrm{S}}\otimes \hat{I}_{\mathrm{E}}\otimes \hat{H}_{\mathrm{B}%
}+\hat{H}_{\mathrm{ES}}\otimes \hat{I}_{\mathrm{B}}+\hat{I}_{\mathrm{E}}%
\otimes \hat{H}_{\mathrm{BS}},
\end{align}
where $\hat{H}_{\mathrm{S}}=h(c_{1}^{\dag}c_{2}+c_{2}^{\dag}c_{1})$, $\hat
{H}_{\mathrm{E}}$, and $\hat{H}_{\mathrm{B}}$ are the Hamiltonians of
\textrm{S}, \textrm{E}, and \textrm{B}, respectively. $\hat{H}_{\mathrm{ES}%
}=\lambda \lbrack a^{\dagger}(c_{1}-ic_{2})+h.c.]$ stands for the coupling
between \textrm{S} and \textrm{E} that leads to a Lindblad operator $\hat
{L}_{\mathrm{ES}}=\sqrt{2\gamma}(c_{1}+ic_{2})\otimes \hat{I}_{\mathrm{B}}$.
$\hat{H}_{\mathrm{BS}}=\gamma_{1}c_{1}^{\dag}c_{1}\otimes \hat{B}_{1}%
+\gamma_{2}c_{2}^{\dag}c_{2}\otimes \hat{B}_{2}$ describes the coupling between
\textrm{S} and \textrm{B}, where $\hat{B}_{1}$ and $\hat{B}_{2}$ are the
operators of the system \textrm{B}. Here, we assume that $\gamma_{1}%
,\gamma_{2}<<\gamma,h.$ In particular, there is \emph{no direct coupling}
between the two environments \textrm{B} and \textrm{E}. In the following
parts, $h$ is set be the unit.\

For the open quantum system \textrm{S} under the Markovian approximation, its
non-unitary dynamics is in general expressed by the quantum master equation in
the Gorini-Kossakowski-Sudarshan-Lindblad (GKSL) form \cite{Lindblad76, GKS76,
BreuerPetruccione}, $\frac{d\rho^{\mathrm{S}}(t)}{dt}\equiv \mathcal{L}%
\rho^{\mathrm{S}}(t)$, where $\mathcal{L}$ is a Liouville super-operator
acting on the (reduced) density matrix $\rho^{\mathrm{S}}$ of the subsystem
\textrm{S}.

In Hermitian systems at finite-T, the quantum statistical mechanics is based
on the thermal equilibrium state. To reach it, one can prepare a unique final
state under time-evolution (from an arbitrary initial state). Using a similar
logic, we introduce a ``\emph{non-Hermitian thermal state}'' (NHTS) which is
the unique final state under time-evolution in a NH system at finite-T (from
an arbitrary initial state). We then give its rigorous mathematical definition
as follows:

\textit{ In thermal }$\mathcal{PT}$\textit{ systems, if the spectrum of
Liouville super-operator }$\mathcal{L}$ \textit{is} $\lambda_{i},$
\textit{i.e.,} $\mathcal{L}\rho_{i}^{\mathrm{S}}=\lambda_{i}\rho
_{i}^{\mathrm{S}},$ \textit{then} \textit{a non-Hermitian thermal state
described by }$\rho_{NHTS}^{\mathrm{S}}$\textit{ is the eigenstate of
}$\mathcal{L}$ \textit{corresponding to the eigenvalue with the maximum real
part }$\lambda_{NHTS}=\max(\operatorname{Re}\lambda_{i})$.

From the above definition, we need to find the possible NHTS by solving the
GKSL equation in order to explore the properties of this thermal\textbf{
}$\mathcal{PT}$ system.

First, we trace out \textrm{E} from the total system and consider its effect
on the combined subsystem \textrm{S+B}. The GKSL equation of the reduced
density matrix $\rho^{\mathrm{S+B}}$ for the combined subsystem \textrm{S+B}
becomes
\begin{align}
\frac{d\rho^{\mathrm{S+B}}}{dt}  &  =-i[\hat{H}_{\mathrm{S+B}},\rho
^{\mathrm{S+B}}]\nonumber \\
&  -\frac{1}{2}\{ \hat{L}_{\mathrm{ES}}^{\dag}\hat{L}_{\mathrm{ES}}%
,\rho^{\mathrm{S+B}}\}+\hat{L}_{\mathrm{ES}}\rho^{\mathrm{S+B}}\hat
{L}_{\mathrm{ES}}^{\dag}.
\end{align}
We make the postselection measurement for the number of particles of the
subsystem \textrm{S} $\left(  c_{1}^{\dag}c_{1}+c_{2}^{\dag}c_{2}\right)
\otimes \hat{I}_{\mathrm{B}}$\textbf{ }so that it is always\textbf{ }$1$
\cite{Aharonov1988}. After these postselection measurements, the quantum
jumping term from $\hat{L}_{\mathrm{ES}}\rho^{\mathrm{S+B}}\hat{L}%
_{\mathrm{ES}}^{\dag}$ will be projected out \cite{Ueda2020}. Then\textrm{
}$\rho^{\mathrm{S+B}}$ satisfies
\begin{equation}
\frac{d\rho^{\mathrm{S+B}}}{dt}=-i(\hat{H}_{\mathrm{S+B},\mathrm{eff}}%
\rho^{\mathrm{S+B}}-\rho^{\mathrm{S+B}}\hat{H}_{\mathrm{S+B},\mathrm{eff}%
}^{\dagger}),
\end{equation}
where $\hat{H}_{\mathrm{S+B},\mathrm{eff}}=\hat{H}_{\mathrm{NH}}\otimes \hat
{I}_{\mathrm{B}}+\hat{I}_{\mathrm{S}}\otimes \hat{H}_{\mathrm{B}}+\hat
{H}_{\mathrm{BS}}$ is the effective Hamiltonian of the combined subsystem
\textrm{S+B} after postselection measurements, and $\hat{H}_{\mathrm{NH}%
}=(h+\gamma)c_{1}^{\dag}c_{2}+(h-\gamma)c_{2}^{\dag}c_{1}-i\gamma$ can
effectively describe the subsystem \textrm{S}. Here, the term $-i\gamma$ in
$\hat{H}_{\mathrm{NH}}$ can be ignored due to postselections
\cite{Yuto2018,Todda2002}.

In short, $\hat{H}_{\mathrm{NH}}$\textbf{ }can be effectively expressed as
$\hat{H}_{\mathrm{NH}}=h\sigma_{x}+i\gamma \sigma_{y}$ on the pseudo-spin space
$(%
\begin{array}
[c]{c}%
\left \vert \uparrow \right \rangle \\
\left \vert \downarrow \right \rangle
\end{array}
)$, where $\left \vert \uparrow \right \rangle $\ and $\left \vert \downarrow
\right \rangle $ denote the quantum states of the single fermion on site $1$
and site $2$, respectively. For $\hat{H}_{\mathrm{NH}}$, at\textbf{ }%
$h=\gamma$\textbf{,} a usual $\mathcal{PT}$-symmetry spontaneous breaking
($\mathcal{PT}$-SSB) occurs: for the case $h>\gamma,$ the energy levels
$\left \vert +\right \rangle $ and $\left \vert -\right \rangle $\textit{ }are
$E_{\pm}=\pm \sqrt{h^{2}-\gamma^{2}}$; for the case $h<\gamma,$ the energy
levels $\left \vert +\right \rangle $ and $\left \vert -\right \rangle $
are\textit{ }$E_{\pm}=\pm i\sqrt{\gamma^{2}-h^{2}}$; and for the case
$h=\gamma,$ the system is at the EP with state coalescing and energy
degeneracy, i.e., $E_{\pm}=0$.

Next, we trace out \textrm{B} from the combined subsystem \textrm{S+B }and
consider it on the NH subsystem \textrm{S.} This issue of \emph{NH open
quantum systems} has never been studied before. We assume that the NH
subsystem \textrm{S} has little influence on the environment \textrm{B} and
the relaxation time of the environment \textrm{B} is much smaller than
\textrm{S}. Besides these conditions, the temperature of a possible NHTS is
also assumed to be the same as that of the thermal environment \textrm{B}.

On one hand, we study the possible NHTS $\rho_{NHTS}^{\mathrm{S}}$ in the
phase with $\mathcal{PT}$-symmetry ($h>\gamma$). Now, the two energy levels of
NH Hamiltonian are purely real, $\pm \sqrt{\gamma^{2}-h^{2}}$. As a result, the
NHTS must be a steady state with the eigenvalue $0$ corresponding to the
Liouville super-operator $\mathcal{L}.$ After solving the GKSL equation,
$\rho_{NHTS}^{\mathrm{S}}$ is obtained as $\frac{h-\gamma}{2h}%
\begin{pmatrix}
\frac{h+\gamma}{h-\gamma} & \sqrt{\frac{h+\gamma}{h-\gamma}}\frac
{1-e^{\omega_{0}\beta_{T}}}{1+e^{\omega_{0}\beta_{T}}}\\
\sqrt{\frac{h+\gamma}{h-\gamma}}\frac{1-e^{\omega_{0}\beta_{T}}}%
{1+e^{\omega_{0}\beta_{T}}} & 1
\end{pmatrix}
.$ More detailed calculations are provided in the Supplementary Materials.

On the other hand, we study the possible NHTS $\rho_{NHTS}^{\mathrm{S}}$ in
the phase with $\mathcal{PT}$-symmetry breaking ($h<\gamma$). Under the
conditions $\gamma_{1},\gamma_{2}<<\gamma,h$, the dissipation term in the
Liouville super-operator only has little effect, i.e. $\mathcal{L}%
\rho^{\mathrm{S}}\approx-i(\hat{H}_{\mathrm{NH}}\rho^{\mathrm{S}}%
-\rho^{\mathrm{S}}\hat{H}_{\mathrm{NH}}^{\dagger}).$ Now, the eigenvalues of
the super-operator $\mathcal{L}$ are obtained as $-2\sqrt{\gamma^{2}-h^{2}},$
$0,$ $0, $ $2\sqrt{\gamma^{2}-h^{2}}$. The eigenstate with the maximum real
part $2\sqrt{\gamma^{2}-h^{2}}$ corresponds to a NHTS\ $\rho_{NHTS}%
^{\mathrm{S}}=|\Psi_{+}^{\mathrm{R}}\rangle \langle \Psi_{+}^{\mathrm{R}}|.$
This can be understood as the fact that the state $|\Psi_{-}^{\mathrm{R}%
}\rangle$ decays much more faster than the state $|\Psi_{+}^{\mathrm{R}%
}\rangle$. Eventually, there is only one eigenstate $|\Psi_{+}^{\mathrm{R}%
}\rangle$ with the largest imaginary part of the eigenvalue $+i\sqrt
{\gamma^{2}-h^{2}}$ left in the system.

Finally, according to the above results from the GKSL equation, we investigate
the properties of NHTSs in the thermal $\mathcal{PT}$ system. These results of
the expected values of $\sigma_{i}$ ($i=x,y,z$), i.e. $n_{i}=\left \langle
\sigma_{i}\right \rangle $, are represented by points on the Bloch sphere in
Fig.2(a) and Fig.2(b). Fig.2(a) indicates the existence of a \emph{unique}
NHTS $\rho_{NHTS}^{\mathrm{S}}$: for $\hat{H}_{\mathrm{NH}}$ with fixed
$\gamma$ ($\gamma=0.2h$), taking the limit $t\rightarrow \infty$, different
initial states will eventually evolve into the same final state $\rho
^{\mathrm{S}}(t\rightarrow \infty)$. That means $\rho^{\mathrm{S}}%
(t\rightarrow \infty)$ is just the NHTS $\rho_{NHTS}^{\mathrm{S}}$; Fig.2(b)
indicates the \emph{abnormality} of the NHTS $\rho_{NHTS}^{\mathrm{S}}$: for
$\hat{H}_{\mathrm{NH}}$ with different values for $\gamma$ ($\gamma=0.3h,$
$\gamma=0.9h$, $\gamma=1.5h$), from the same initial state, for example, a
fully thermalized state described by $\rho^{\mathrm{S}}(t=0)=\frac{1}%
{2}I_{2\times2}$, the system will eventually evolve into different NHTSs
$\rho_{NHTS}^{\mathrm{S}}(\gamma)$.

\begin{figure}[ptb]
\includegraphics[clip,width=0.48\textwidth]{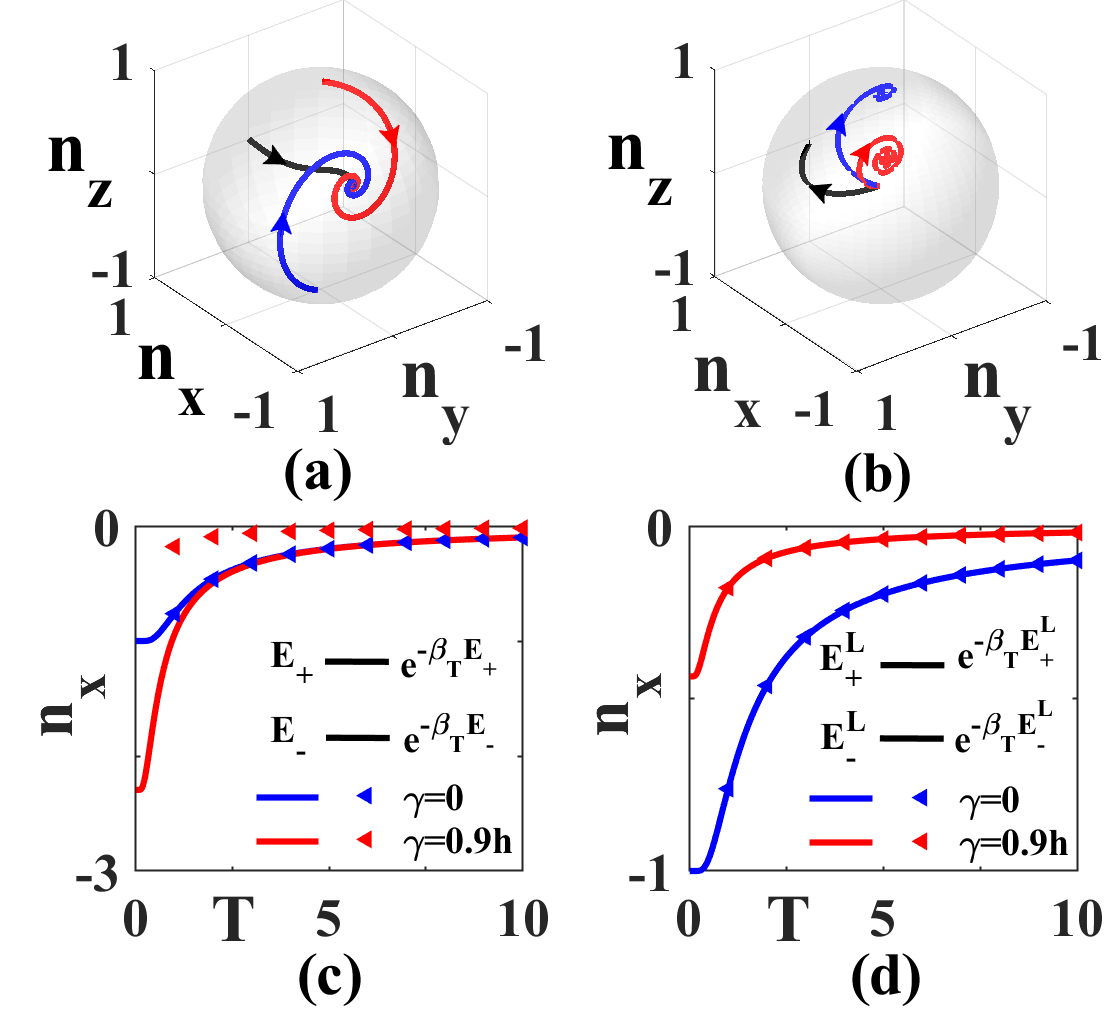}\caption{(a)
Time-evolution of non-Hermitian (NH) systems for fixed $\gamma$ ($\gamma
=0.2h$) from the different initial states at temperature $k_{B}T=h$, where
$\gamma_{0}=0.05h$; (b) Time-evolution of NH systems for different $\gamma$
($\gamma=0.3h,$ $\gamma=0.9h$, $\gamma=1.5h$) from the same fully thermalized
initial state, where $\gamma_{0}=0.05h$ and $k_{B}T=10h$. The arrows denote
the direction of time-evolution; (c) The comparison between the results from
usual Boltzmann distribution (solid lines) and those obtained by directly
solving the Gorini-Kossakowski-Sudarshan-Lindblad (GKSL) equation (the small
triangles); (d) The comparison between the results from the
Liouvillian-Boltzmann distribution (solid lines) and those derived by directly
solving the GKSL equation (the small triangles).}%
\end{figure}

Let us\textbf{ }check the validity of the BZ law in this thermal
$\mathcal{PT}$ system.

For the Hermitian case ($\gamma=0$), different thermal equilibrium states obey
the BZ law, i.e., $P_{+}^{T}=e^{-\beta_{T}E_{+}}$ and $P_{-}^{T}=e^{-\beta
_{T}E_{-}}$, where $E_{\pm}=\pm h$\ are energy levels. In Fig.2(c), the blue
line and blue triangles represent these expected values of $\sigma_{x}$
($n_{x}$) from the BZ law and those derived by directly solving the GKSL
equation, respectively. The consistence between the results from the BZ law
and those from the GKSL equation verifies the correctness of both approaches.

Then we study the NH case. For NHTSs with $\gamma=0.9h$, we can also derive
$n_{x}$ by using the BZ law, i.e. $n_{x}=\frac{1}{e^{-\beta_{T}E_{-}}+e^{-\beta
_{T}E_{+}}}\mathrm{Tr}(\sigma_{x}\cdot e^{-\beta_{T}\hat{H}_{\mathrm{NH}}}).$
In Fig.2(c), the red line and red triangles represent $n_{x}$ from the BZ law
and those obtained by directly solving the GKSL equation, respectively. One
can see that the results derived by means of the two mentioned approaches are
noticeably contradictory in this case, which means that the BZ law does not
hold in this situation. Consequently, one can safely conclude that the usual
BZ law for thermal equilibrium states in Hermitian systems does not work anymore in NH
systems! \emph{The immediate questions would be how to understand this
violation of the BZ law in NHTSs, and whether there exists a new law that
explains these abnormal results for a thermal $\mathcal{PT}$ system}.

Before developing a systematic theory, we provide a physical explanation. In
fact, NHTSs come from NH systems rather than from free ones. To realize a NH
system, we controlled the open thermal $\mathcal{PT}$ system by performing a
postselection and projecting out the quantum jumping term. The postselection
measurement would result in a continuous information backflow from the NH
environment \textrm{E} to the subsystem \textrm{S} \cite{Ueda17}. These
control actions break down the equal probability of different microscopic
states and lead to a new distribution law.

In order to better describe this new distribution law, we develop a quantum
Liouvillian statistics theory to systematically and analytically characterize
the thermal $\mathcal{PT}$ systems. In particular, it is a new type of
distribution, that we call \emph{Liouvillian-Boltzmann distribution (LBZ)}
that governs the distribution of NHTSs at finite-T.

We focus on the case with real energy levels ($h>\gamma$). By doing a
similarity transformation (ST) $\mathcal{\hat{S}}=e^{\beta_{\mathrm{NH}}%
\hat{H}^{\prime}},$ $\hat{H}_{\mathrm{NH}}$ can be transformed into a
Hermitian Hamiltonian $\hat{H}_{0}=\sqrt{h^{2}-\gamma^{2}}\sigma_{x}$, i.e.,
$\hat{H}_{\mathrm{NH}}=\mathcal{\hat{S}}\hat{H}_{0}\mathcal{\hat{S}}^{-1}.$
Here, the real number $\beta_{\mathrm{NH}}=\frac{1}{2}\ln \frac{h+\gamma
}{h-\gamma}$ characterizes the strength of the NH terms and the Hermitian
operator $\hat{H}^{\prime}=\frac{1}{2}\sigma_{z}$ determines the form of the
NH terms. Under the NH TS $\mathcal{\hat{S}}$, the energy levels\textit{
}$E_{\pm}$ of $\hat{H}_{\mathrm{NH}}$ are same to those of the Hermitian model
$\hat{H}_{0}.$ The corresponding eigenstates of $\hat{H}_{\mathrm{NH}}$ become
$|\Psi_{n}^{\mathrm{R}}\rangle=\mathcal{\hat{S}}|\Psi_{0,n}\rangle$, where
$n=\pm$ and $|{\Psi}_{0,n}\rangle$ denotes the normalized eigenstates of
$\hat{H}_{0}$. It is obvious that the effect from the NH terms leads to the
additional NH TS $\mathcal{\hat{S}}$ that breaks down the equal probability of
different microscopic states.

Based on these physical quantities, we define the \emph{analytical }formula of
the density matrix $\rho_{NHTS}^{\mathrm{S}}$.

\textit{The density matrix for the thermal state of a NH system at temperature
}$T=\frac{1}{k_{B}\beta_{T}}$\textit{ described by }$\hat{H}_{\mathrm{NH}%
}=\mathcal{\hat{S}}\hat{H}_{0}\mathcal{\hat{S}}^{-1}$\textit{ (}%
$\mathcal{\hat{S}}=e^{\beta_{\mathrm{NH}}\hat{H}^{\prime}}$\textit{) is given
as }%
\begin{equation}
\label{eq1}\rho_{NHTS}^{\mathrm{S}}=\sum_{n=\pm}|\Psi_{n}^{\mathrm{R}}\rangle
e^{-\beta_{T}E_{n}}\langle \Psi_{n}^{\mathrm{R}}|=\mathcal{\hat{S}}\rho
_{0}\mathcal{\hat{S}}^{\dagger},
\end{equation}
\textit{where }$\rho_{0}$\textit{ is the density matrix for Hermitian
model}\emph{ }$\hat{H}_{0}$\textit{. }In \cite{com}, we provide the reason to
write down the above equation \eqref{eq1}.

To simplify the Liouvillian physics for NHTSs, we introduce an effective
Hamiltonian $\hat{H}_{\mathrm{L}}$ as $e^{-\beta_{T}\hat{H}_{\mathrm{L}}%
}=\mathcal{\hat{S}}(e^{-\beta_{T}\hat{H}_{0}})\mathcal{\hat{S}}^{\dagger}$ or
\begin{equation}
\hat{H}_{\mathrm{L}}=-\frac{1}{\beta_{T}}\ln[\mathcal{\hat{S}}(e^{-\beta
_{T}\hat{H}_{0}})\mathcal{\hat{S}}^{\dagger}].
\end{equation}
In this paper, we call the temperature-dependent Hamiltonian $\hat
{H}_{\mathrm{L}}$ to\ be \emph{Liouvillian Hamiltonian} (LH). Its
corresponding eigenvalues $E_{n}^{\mathrm{L}}$ and eigenstates are called
\emph{Liouvillian energy levels} and \emph{Liouvillian states}, respectively.
According to $\hat{H}_{\mathrm{L}}=\hat{H}_{\mathrm{L}}^{\dagger}$,
$E_{n}^{\mathrm{L}}$ must be real.

For NHTSs in this NH system, although the energy levels do not change under
STs, the weights of NHTSs are deformed. Consequently, the usual BZ law
$P_{n}^{T}=\frac{1}{Z_{0}}e^{-\beta_{T}E_{n}}$ is replaced by the \emph{
Liouvillian-Boltzmann distribution }(LBZ) law \cite{gar,ma}, i.e.,
\begin{equation}
P_{n}^{T}=\frac{1}{Z_{0}}e^{-\beta_{T}E_{n}}\rightarrow P_{n}^{T}=\frac
{1}{Z_{NHTS}^{\mathrm{S}}}e^{-\beta_{T}E_{n}^{\mathrm{L}}},
\end{equation}
where $E_{n}^{\mathrm{L}}$ is a Liouvillian energy level rather than the
energy level $E_{n}$ and $Z_{NHTS}^{\mathrm{S}}=Tr(\rho_{NHTS}^{\mathrm{S}})$
are the partition functions for the NHTS. To distinguish the phenomenon of
weights in the thermal $\mathcal{PT}$ system from those in Hermitian thermal
systems, we call the corresponding weight $P_{n}^{T}$ \emph{the Liouvillian
weight}.

In particular, when the temperature is high, $\beta_{T}\rightarrow0$, the
density matrix $\rho_{NHTS}^{\mathrm{S}}=e^{-\beta_{T}\hat{H}_{\mathrm{L}}}$
for a NH system with real spectra is reduced to $\rho_{NHTS}^{\mathrm{S}}%
\sim \mathcal{\hat{S}\hat{S}}^{\dagger}=e^{2\beta_{\mathrm{NH}}\hat{H}^{\prime
}}$, and the Liouvillian weight turns into $P_{n}^{T}=\frac{1}{Z_{NHTS}%
^{\mathrm{S}}}e^{-\beta_{\mathrm{NH}}(-2E_{n}^{\prime})}$, where
$E_{n}^{\prime}$ is the ``energy levels'' of $\hat{H}^{\prime}$.

By employing the quantum Liouvillian statistical theory, we study
thermodynamic properties for the NHTSs $\rho_{NHTS}^{\mathrm{S}}$ of the
thermal $\mathcal{PT}$ system.

\begin{figure}[ptb]
\includegraphics[clip,width=0.47\textwidth]{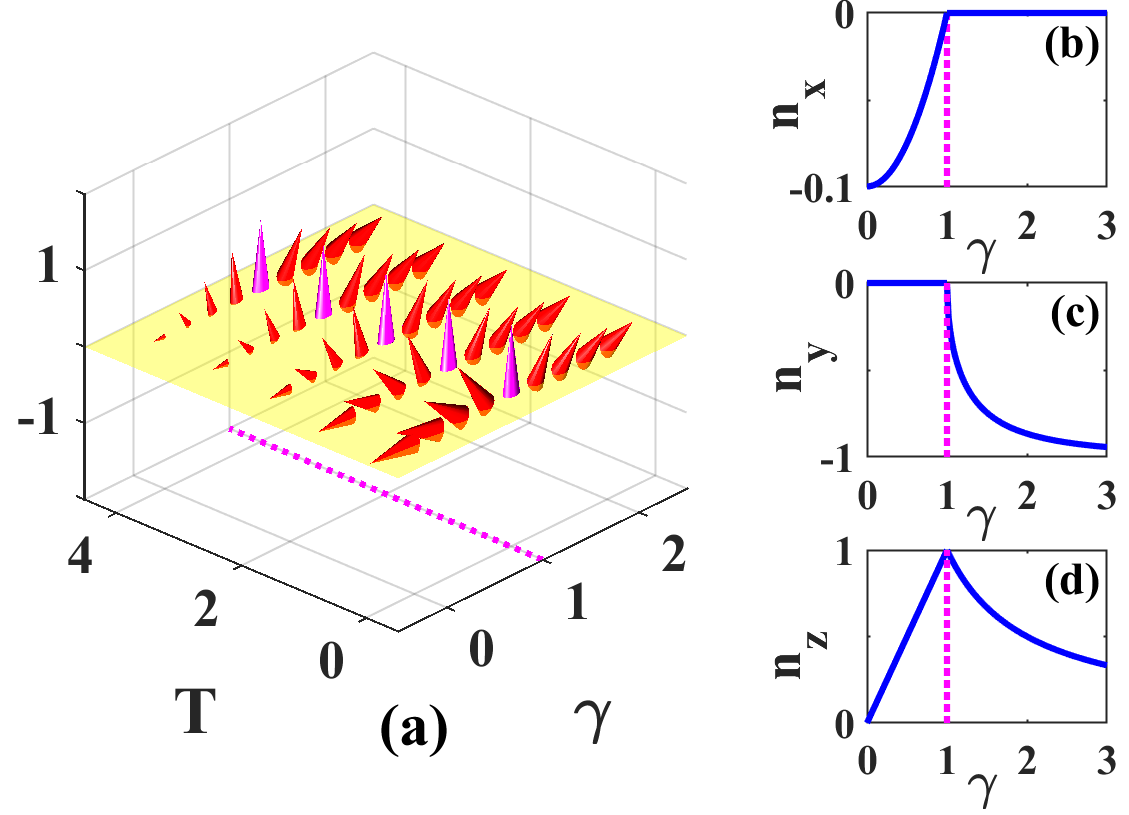}\caption{(a) Expected
values of the spin operator $n_{i}=\left \langle \sigma_{i}\right \rangle $ from
Liouvillian-Boltzmann distribution for the non-Hermitian thermal states
$\rho_{NHTS}^{\mathrm{S}}$ of the thermal $\mathcal{PT}$ system. (b), (c), and
(d) depict $n_{x}$, $n_{y}$, $n_{z}$ at temperature $k_{B}T=10h$,
respectively. }%
\end{figure}

These physical properties of the original NH system $\hat{H}_{\mathrm{NH}}$ at
finite-T correspond to those of a Hermitian system of LH, $\hat{H}%
_{\mathrm{L}}$. According to $e^{-\beta_{T}\hat{H}_{\mathrm{L}}}%
=\mathcal{\hat{S}}(e^{-\beta_{T}\hat{H}_{0}})\mathcal{\hat{S}}^{\dagger},$ we
obtain the analytical result as $e^{-\beta_{T}\hat{H}_{\mathrm{L}}}=\sigma
_{x}A+\sigma_{z}B+\mathrm{I}C$, where $\mathrm{I}=(%
\begin{array}
[c]{cc}%
1 & 0\\
0 & 1
\end{array}
),$ $A=-\sinh(\beta_{T}\sqrt{h^{2}-\gamma^{2}}),$ $B=\sinh(\beta_{\mathrm{NH}%
})\cdot \cosh(\beta_{T}\sqrt{h^{2}-\gamma^{2}})$ and $C=\cosh(\beta
_{\mathrm{NH}})\cdot \cosh(\beta_{T}\sqrt{h^{2}-\gamma^{2}})$. As a result, the
LH is derived as
\[
\hat{H}_{\mathrm{L}}=-\frac{1}{\beta_{T}}\frac{\cosh^{-1}C}{\left \vert \vec
{r}_{\mathrm{eff}}\right \vert }(\vec{\sigma}\cdot \vec{r}_{\mathrm{eff}}),
\]
where $\vec{\sigma}=(\sigma_{x},\sigma_{y},\sigma_{z})$, $\vec{r}%
_{\mathrm{eff}}=(A,0,B)$ and $\left \vert \vec{r}_{\mathrm{eff}}\right \vert
=\sqrt{A^{2}+B^{2}}$. Hence, the Liouvillian energy levels\textbf{ }(the
eigenvalues of $\hat{H}_{\mathrm{L}}$) are
\[
E_{\pm}^{\mathrm{L}}=\pm \frac{1}{\beta_{T}}\cosh^{-1}\left(  \cosh
\beta_{\mathrm{NH}}\cdot \cosh(\beta_{T}/2)\right)  ,
\]
that are quite different from the energy levels (the eigenvalues for $\hat
{H}_{\mathrm{NH}}$) $E_{\pm}=\pm \sqrt{h^{2}-\gamma^{2}}$. According to LBZ
law, we have
\[
P_{n}^{T}=\frac{1}{Z_{NHTS}^{\mathrm{S}}}\exp[\frac{\cosh^{-1}C}{\left \vert
\vec{r}_{\mathrm{eff}}\right \vert }(\vec{\sigma}\cdot \vec{r}_{\mathrm{eff}%
})].
\]
The expected values of $\sigma_{i}$ ($i=x,y,z$) are defined as $n_{i}%
=\left \langle \sigma_{i}\right \rangle =\frac{1}{Z_{NHTS}^{\mathrm{S}}%
}\mathrm{Tr}(\sigma_{i}\cdot e^{-\beta_{T}\hat{H}_{\mathrm{L}}}).$

After straightforward calculations, we have $\vec{n}=(n_{x},n_{y}%
,n_{z})=(\frac{A}{C},0,\frac{B}{C})$ in the region of $h>\gamma$. We then
check the validity of the LBZ law by comparing its results ($n_{x}=\frac{A}%
{C}$) with those from solving the GKSL equation. As shown in Fig.2(d), they
have the exact consistence of those derived by directly solving the GKSL
equation (small triangles). Fig.3(a) shows the results of $\vec{n}$. From it,
one can see that when $\beta_{\mathrm{NH}}$ increases from zero to infinite,
the spin direction changes from the $x$-direction to the $z$-direction.
Fig.3(b)-(d) show the behavior of $n_{i}=\left \langle \sigma_{i}\right \rangle
$ at $k_{B}T=10h$, respectively. Approaching to EP $h\rightarrow \gamma$ (or
$\beta_{\mathrm{NH}}\rightarrow \infty$), we have $\hat{H}_{\mathrm{L}%
}\rightarrow \frac{\beta_{\mathrm{NH}}}{\beta_{T}}\sigma_{z}$ at finite-T. The
external field $\left \vert \frac{\beta_{\mathrm{NH}}}{\beta_{T}}\right \vert $
in LH $\hat{H}_{\mathrm{L}}$ becomes \emph{infinite}. As a result, $\vec{n}$
is fixed to be polarized along the $z$-direction with saturated amplitude $1$
(See the magenta arrows in Fig.2(a)) marked by a magenta line). For the region
of\textbf{ }$h<\gamma$\textbf{,} the NHTS becomes a pure state described by
$|\Psi_{+}^{\mathrm{R}}\rangle$ and we have $\vec{n}=(0,-\sqrt{1-(\frac
{h}{\gamma})^{2}},\frac{h}{\gamma})$. As shown in Fig.2(a), the red arrows
(the expected values of $\sigma_{i}$) swerve from the $z$-direction to the
$y$-direction with increasing\textbf{ }$\gamma$.

Finally, we discuss the possible thermodynamic phase transition at
the EP $\ h=\gamma$.

From the above discussions we concluded that the average spin operator $n_{i}$
is always continuous when crossing over the EP. Furthermore, we calculate the
derivatives of spin average values $\frac{\partial \vec{n}}{\partial \gamma}$.
Because $h=\gamma$ is a transition from real spectra to complex, the NHTSs of
both sides ($h>\gamma$\ and $h<\gamma$) are described by different functions.
Thus, $\frac{\partial \vec{n}}{\partial \gamma}$ becomes discontinuous at
$h=\gamma$. The discontinuity of $\frac{\partial \vec{n}}{\partial \gamma}$
indicates a ``continuous'' thermodynamic phase transition. In particular,
there exists a $\nu=\frac{1}{2}$ critical rule for $\frac{\partial \vec{n}%
}{\partial \gamma}$: at finite-T $T\neq0$, $\frac{\partial \vec{n}}%
{\partial \gamma}=(2\beta_{T},0,h^{-1})$\ for $h\rightarrow \gamma+0^{+}$ and
$\frac{\partial \vec{n}}{\partial \gamma}=(0,-(2h)^{^{-\frac{1}{2}}}%
(\gamma-h)^{-\nu},-\gamma^{-1})$\ for $\gamma \rightarrow h+0^{+}$; at zero
temperature $T=0$, $\frac{\partial \vec{n}}{\partial \gamma}=((2h)^{-\frac{1}%
{2}}(h-\gamma)^{-\nu},0,h^{-1})$\ for $h\rightarrow \gamma+0^{+}$ and
$\frac{\partial \vec{n}}{\partial \gamma}=(0,-(2h)^{-\frac{1}{2}}(\gamma
-h)^{-\nu},-\gamma^{-1})$\ for $\gamma \rightarrow h+0^{+}$. To emphasize the
strangeness of the $\nu=\frac{1}{2}$ critical rule, we call it \emph{zero
temperature anomaly} for the thermodynamic phase transition at EP. See the
detailed calculations in the Supplementary Materials.

In this paper we studied the physics of thermal $\mathcal{PT}$ systems. Our
results show that due to the postselection measurements, for the thermal
$\mathcal{PT}$ system both the usual BZ law and Equal Probability Principle do
not hold anymore. To characterize this abnormal behavior in NH systems at
finite-T, a quantum Liouvillian statistics theory was developed, where the
usual BZ law $P_{n}^{T}\sim e^{-E_{n}/k_{B}T}$ is replaced by the LBZ law
$P_{n}^{T}\sim e^{-E_{n}^{\mathrm{L}}/k_{B}T}$, where $E_{n}^{\mathrm{L}}$ is
a Liouvillian energy level rather than an energy level $E_{n}$. Based on the
new theory, analytical results of thermodynamic properties for the thermal
$\mathcal{PT}$ system were derived. We found that a ``continuous''
thermodynamic phase transition occurs at the EP, where there exists a
zero-temperature anomaly. In the future, we plan to study the thermal states
of more complex NH models and to explore the possible exotic phenomena in NH
systems at finite-T.

\acknowledgments This work was supported by NSFC Grant No. 11974053, 12174030.
We are grateful to Wei Yi, Shu Chen, Gao-Yong Sun, Lin-Hu Li, Yi-Bin Guo,
Xue-Xi Yi, and Ching-Hua Lee for helpful discussions that contributed to
clarifying some aspects related to the present work.\

\end{document}


\title{Supplemental Material for \textquotedblleft Physics of PT-Symmetric Quantum
Systems at Finite Temperature"}
\author{Qian Du}
\thanks{Who has same contribution to this work}
\affiliation{Center for Advanced Quantum Studies, Department of Physics, Beijing Normal
University, Beijing 100875, China}
\author{Kui Cao}
\thanks{Who has same contribution to this work}
\affiliation{Center for Advanced Quantum Studies, Department of Physics, Beijing Normal
University, Beijing 100875, China}
\author{Su-Peng Kou}
\thanks{Corresponding author}
\email{spkou@bnu.edu.cn}
\affiliation{Center for Advanced Quantum Studies, Department of Physics, Beijing Normal
University, Beijing 100875, China}
\maketitle
\tableofcontents

\section{The detailed calculations of the density matrix for the non-Hermitian
thermal state in the phase with $\mathcal{PT}$-symmetry ($h>\gamma$)}

In this section, we directly solve Gorini-Kossakowski-Sudarshan-Lindblad
(GKSL) equation and obtain the density matrix $\rho_{NHTS}^{\mathrm{S}}$ for a
non-Hermitian thermal state (NHTS) in the phase with $\mathcal{PT}$-symmetry
($h>\gamma$). The followings are detailed calculations.

Firstly, we derive an effective non-Hermitian (NH) Hamiltonian after the
postselection measurements based on the GKSL equation.

For the total system \textrm{S+B+E}, we trace out \textrm{E} from it and get
the GKSL equation of the reduced density matrix $\rho^{\mathrm{S+B}}$
\begin{align}
\frac{d\rho^{\mathrm{S+B}}}{dt}  &  =-i[\hat{H}_{\mathrm{S+B}},\rho
^{\mathrm{S+B}}]\nonumber \\
&  -\frac{1}{2}\{ \hat{L}_{\mathrm{ES}}^{\dag}\hat{L}_{\mathrm{ES}}%
,\rho^{\mathrm{S+B}}\}+\hat{L}_{\mathrm{ES}}\rho^{\mathrm{S+B}}\hat
{L}_{\mathrm{ES}}^{\dag},
\end{align}
where the Lindblad operator is $\hat{L}_{\mathrm{ES}}=\sqrt{2\gamma}%
(c_{1}+ic_{2})\otimes \hat{I}_{\mathrm{B}}$.\textrm{ }We make the postselection
measurement for the number of particles of the subsystem \textrm{S}\textbf{
}$\left(  c_{1}^{\dag}c_{1}+c_{2}^{\dag}c_{2}\right)  \otimes \hat
{I}_{\mathrm{B}}$\textbf{ }so that it is always\textbf{ }$1$. After the
postselection measurements, the quantum jumping term $\hat{L}_{\mathrm{ES}%
}\rho^{\mathrm{S+B}}\hat{L}_{\mathrm{ES}}^{\dag}$ will be projected out. Then
$\rho_{\mathrm{S+B}}$ becomes
\begin{equation}
\frac{d\rho^{\mathrm{S+B}}}{dt}=-i(\hat{H}_{\mathrm{S+B},\mathrm{eff}}%
\rho^{\mathrm{S+B}}-\rho^{\mathrm{S+B}}\hat{H}_{\mathrm{S+B},\mathrm{eff}%
}^{\dagger}), \label{eq2}%
\end{equation}
where
\[
\hat{H}_{\mathrm{S+B},\mathrm{eff}}=\hat{H}_{\mathrm{NH}}\otimes \hat
{I}_{\mathrm{B}}+\hat{I}_{\mathrm{S}}\otimes \hat{H}_{\mathrm{B}}+\hat
{H}_{\mathrm{BS}}%
\]
with
\[
\hat{H}_{\mathrm{NH}}=(h+\gamma)c_{1}^{\dag}c_{2}+(h-\gamma)c_{2}^{\dag}%
c_{1}-i\gamma \left(  c_{1}^{\dag}c_{1}+c_{2}^{\dag}c_{2}\right)
\]
and%
\[
\hat{H}_{\mathrm{BS}}=\gamma_{1}c_{1}^{\dag}c_{1}\otimes \hat{B}_{1}+\gamma
_{2}c_{2}^{\dag}c_{2}\otimes \hat{B}_{2}.
\]
Because we have fixed the number of particles in subsystem \textrm{S} to $1$
by postselection, the third term of\textbf{ }$\hat{H}_{\mathrm{NH}}$ become a
constant $-i\gamma$. During the postselection measurements, $\rho
^{\mathrm{S+B}}$ at each moment needs to be normalized and thus\textbf{ }we
can ignore the term $-i\gamma$.

When considering the postselection process, the subsystem \textrm{S} can be
effectively described by an effective NH Hamiltonian
\begin{equation}
\hat{H}_{\mathrm{NH}}=(h+\gamma)c_{1}^{\dag}c_{2}+(h-\gamma)c_{2}^{\dag}c_{1}.
\end{equation}
In the case of $h>\gamma$, its eigenvalues are $E_{-}=-\sqrt{h^{2}-\gamma^{2}%
}$ with the right eigenstate $|-\rangle_{R}=\frac{1}{\sqrt{2}}\left(
\begin{array}
[c]{cc}%
1 & y
\end{array}
\right)  ^{T}$ and the left eigenstate $|-\rangle_{L}=\frac{1}{\sqrt{2}%
}\left(
\begin{array}
[c]{cc}%
1 & x
\end{array}
\right)  ^{T}$, $E_{+}=\sqrt{h^{2}-\gamma^{2}}$ with the right eigenstate
$|+\rangle_{R}=\frac{1}{\sqrt{2}}\left(
\begin{array}
[c]{cc}%
1 & -y
\end{array}
\right)  ^{T}$ and the left eigenstate $|+\rangle_{L}=\frac{1}{\sqrt{2}%
}\left(
\begin{array}
[c]{cc}%
1 & -x
\end{array}
\right)  $, respectively. Here, we have $x=\sqrt{\frac{h+\gamma}{h-\gamma}}$,
$y=1/x$.

Secondly, we calculate the Liouville super-operator $\mathcal{L}$ under the
Born-Markov approximation\textbf{ }in the interaction picture for the NH open
quantum system, which\textbf{ }has never been studied before.\textbf{ }

In general, the derivation of a quantum Markovian master equation is performed
in the interaction picture. Thus, we write Eq.~\eqref{eq2} as in the
interaction picture
\begin{equation}
\frac{d}{dt}\rho_{I}^{\mathrm{S+B}}(t)=-i(\hat{V}_{I}\left(  t\right)
\rho_{I}^{\mathrm{S+B}}(t)-\rho_{I}^{\mathrm{S+B}}(t)\hat{V}_{I}^{\dag}\left(
t\right)  ), \label{eq4}%
\end{equation}
where $\rho_{I}^{\mathrm{S+B}}(t)=e^{iH_{\mathrm{eff},0}t}\rho^{\mathrm{S+B}%
}(t)e^{-iH_{\mathrm{eff},0}^{\dag}t}$, $\hat{V}_{I}\left(  t\right)
=e^{iH_{\mathrm{eff},0}t}\hat{H}_{\mathrm{BS}}e^{-iH_{\mathrm{eff},0}t}$ and
$\hat{H}_{\mathrm{eff},0}=\hat{H}_{\mathrm{NH}}\otimes \hat{I}_{\mathrm{B}%
}+\hat{I}_{\mathrm{S}}\otimes \hat{H}_{\mathrm{B}}$. Its equivalent integral
form is
\begin{equation}
\rho_{I}^{\mathrm{S+B}}(t)=\rho_{I}^{\mathrm{S+B}}(0)-i\int_{0}^{t}ds\left[
\hat{V}_{I}\left(  s\right)  \rho_{I}^{\mathrm{S+B}}\left(  s\right)
-\rho_{I}^{\mathrm{S+B}}\left(  s\right)  \hat{V}_{I}^{\dag}\left(  s\right)
\right]  . \label{eq5}%
\end{equation}
We substitute Eq.~\eqref{eq5} into Eq.~\eqref{eq4} and obtain
\begin{align}
\frac{d}{dt}\rho_{I}^{\mathrm{S+B}}(t)  &  =-i\left[  \hat{V}_{I}\left(
t\right)  \rho_{I}^{\mathrm{S+B}}(0)-\rho_{I}^{\mathrm{S+B}}(0)\hat{V}%
_{I}^{\dag}\left(  t\right)  \right] \nonumber \\
&  -\bigg \{ \hat{V}_{I}\left(  t\right)  \int_{0}^{t}ds\left[  \hat{V}%
_{I}\left(  s\right)  \rho_{I}^{\mathrm{S+B}}\left(  s\right)  -\rho
_{I}^{\mathrm{S+B}}\left(  s\right)  \hat{V}_{I}^{\dag}\left(  s\right)
\right] \nonumber \\
&  -\int_{0}^{t}ds\left[  \hat{V}_{I}\left(  s\right)  \rho_{I}^{\mathrm{S+B}%
}\left(  s\right)  -\rho_{I}^{\mathrm{S+B}}\left(  s\right)  \hat{V}_{I}%
^{\dag}\left(  s\right)  \right]  \hat{V}_{I}^{\dag}\left(  t\right)
\bigg \}.
\end{align}
Taking the partial trace over the degrees of freedom of the environment
\textrm{B }for the above equation, we give the reduced density matrix
$\rho_{I}^{\mathrm{S}}(t)$ in the subsystem \textrm{S }
\begin{align}
\frac{d}{dt}\rho_{I}^{\mathrm{S}}(t)  &  =-itr_{\mathrm{B}}\left[  \hat{V}%
_{I}\left(  t\right)  \rho_{I}^{\mathrm{S+B}}(0)-\rho_{I}^{\mathrm{S+B}%
}(0)\hat{V}_{I}^{\dag}\left(  t\right)  \right] \nonumber \\
&  -tr_{\mathrm{B}}\bigg \{ \hat{V}_{I}\left(  t\right)  \int_{0}^{t}ds\left[
\hat{V}_{I}\left(  s\right)  \rho_{I}^{\mathrm{S+B}}\left(  s\right)
-\rho_{I}^{\mathrm{S+B}}\left(  s\right)  \hat{V}_{I}^{\dag}\left(  s\right)
\right] \nonumber \\
&  -\int_{0}^{t}ds\left[  \hat{V}_{I}\left(  s\right)  \rho_{I}^{\mathrm{S+B}%
}\left(  s\right)  -\rho_{I}^{\mathrm{S+B}}\left(  s\right)  \hat{V}_{I}%
^{\dag}\left(  s\right)  \right]  \hat{V}_{I}^{\dag}\left(  t\right)
\bigg \}.
\end{align}
Here, we assume \cite{open}
\begin{equation}
tr_{\mathrm{B}}\left[  \hat{V}_{I}\left(  t\right)  \rho_{I}^{\mathrm{S+B}%
}(0)-\rho_{I}^{\mathrm{S+B}}(0)\hat{V}_{I}^{\dag}\left(  t\right)  \right]
=0.
\end{equation}
Additionally, we use the Born approximation that the subsystem \textrm{S} has
little influence on the environment \textrm{B}, so $\rho_{I}^{\mathrm{S+B}%
}(s)\sim \rho_{I}^{\mathrm{S}}(s)\otimes \rho_{I}^{\mathrm{B}}$. By the Markov
approximation $\rho_{I}^{\mathrm{S}}(s)\sim \rho_{I}^{\mathrm{S}}(t)$, we have
\begin{align}
\frac{d}{dt}\rho_{I}^{\mathrm{S}}(t)  &  =-tr_{\mathrm{B}}\bigg \{ \hat{V}%
_{I}\left(  t\right)  \int_{0}^{t}ds\left[  \hat{V}_{I}\left(  s\right)
\left(  \rho_{I}^{\mathrm{S}}(t)\otimes \rho_{I}^{\mathrm{B}}\right)  -\left(
\rho_{I}^{\mathrm{S}}(t)\otimes \rho_{I}^{\mathrm{B}}\right)  \hat{V}_{I}%
^{\dag}\left(  s\right)  \right] \nonumber \\
&  -\int_{0}^{t}ds\left[  \hat{V}_{I}\left(  s\right)  \left(  \rho
_{I}^{\mathrm{S}}(t)\otimes \rho_{I}^{\mathrm{B}}\right)  -\left(  \rho
_{I}^{\mathrm{S}}(t)\otimes \rho_{I}^{\mathrm{B}}\right)  \hat{V}_{I}^{\dag
}\left(  s\right)  \right]  \hat{V}_{I}^{\dag}\left(  t\right)  \bigg \}.
\end{align}
We substitute $s$ by $t-s$, the above equation can be expressed as%
\begin{equation}
\frac{d}{dt}\rho_{I}^{\mathrm{S}}(t)=tr_{\mathrm{B}}\int_{0}^{t}ds[\hat{V}%
_{I}\left(  t-s\right)  \rho_{I}^{\mathrm{S}}(t)\otimes \rho_{I}^{\mathrm{B}%
}\hat{V}_{I}^{\dag}\left(  t\right)  -\hat{V}_{I}\left(  t\right)  \hat{V}%
_{I}\left(  t-s\right)  \rho_{I}^{\mathrm{S}}(t)\otimes \rho_{I}^{\mathrm{B}%
}]+h.c.. \label{eq10}%
\end{equation}

Using eigenstates of the $\hat{H}_{\mathrm{NH}}$ in $h>\gamma$, we insert the
identity operator $\hat{I}=\sum_{m=+-}\left \vert m\right \rangle _{R}%
\left \langle m\right \vert _{L}\otimes \hat{I}_{B}$ in $\hat{V}_{I}\left(
t\right)  =e^{i\hat{H}_{\mathrm{eff},0}t}\hat{H}_{\mathrm{BS}}e^{-i\hat
{H}_{\mathrm{eff},0}t}$ and get%
\begin{align}
\hat{V}_{I}\left(  t\right)   &  =e^{i\hat{H}_{\mathrm{eff},0}t}\left[
\sum_{a}\sum_{m=+,-}\left \vert m\right \rangle _{R}\left \langle m\right \vert
_{L}\gamma_{a}c_{a}^{\dag}c_{a}\sum_{n=+,-}\left \vert n\right \rangle
_{R}\left \langle n\right \vert _{L}\otimes \hat{B}_{a}\right]  e^{-i\hat
{H}_{\mathrm{eff},0}t}\nonumber \\
&  =\sum_{a}\sum_{m=+,-}\sum_{n}e^{iE_{m}t}\left \vert m\right \rangle
_{R}\left \langle m\right \vert _{L}\gamma_{a}c_{a}^{\dag}c_{a}\left \vert
n\right \rangle _{R}\left \langle n\right \vert _{L}e^{-iE_{n}t}\otimes \hat
{B}_{a}\left(  t\right) \nonumber \\
&  =\sum_{a}\sum_{m=+,-}\sum_{\omega}e^{-i\omega t}\left \vert m\right \rangle
_{R}\left \langle m\right \vert _{L}\gamma_{a}c_{a}^{\dag}c_{a}\left \vert
m+\omega \right \rangle _{R}\left \langle m+\omega \right \vert _{L}\otimes \hat
{B}_{a}\left(  t\right)  ,
\end{align}
where $\omega=E_{n}-E_{m}$ is real and $\left \vert m+\omega \right \rangle $
means the state with energy $E_{n}=E_{m}+\omega$. Here we define an operator
$\hat{A}_{a}(\omega)$,%
\begin{equation}
\hat{A}_{a}(\omega)=\sum_{m=+-}\left \vert m\right \rangle _{R}\left \langle
m\right \vert _{L}\gamma_{a}c_{a}^{\dag}c_{a}\left \vert m+\omega \right \rangle
_{R}\left \langle m+\omega \right \vert _{L}%
\end{equation}
with $a=1$, $2$.

Now, $\hat{V}_{I}\left(  t\right)  $ is written as
\begin{equation}
\hat{V}_{I}\left(  t\right)  =\sum_{a}\sum_{\omega}e^{-i\omega t}\hat{A}%
_{a}(\omega)\otimes \hat{B}_{a}(t).
\end{equation}
Substituting this form of $\hat{V}_{I}(t)$ to Eq.~\eqref{eq10}, we get%
\begin{align}
\label{eq14}\frac{d}{dt}\rho_{I}^{\mathrm{S}}(t)  &  =\sum_{b}\sum_{\omega
}\sum_{a}\sum_{\omega_{1}}e^{i\left(  \omega_{1}-\omega \right)  t}\hat{A}%
_{b}(\omega)\rho_{I}^{\mathrm{S}}(t)\hat{A}_{a}^{\dag}(\omega_{1})\int_{0}%
^{t}dse^{i\omega s}tr_{\mathrm{B}}\left(  \hat{B}_{b}(t-s)\rho_{I}%
^{\mathrm{B}}\hat{B}_{a}^{\dag}(t)\right) \nonumber \\
&  -\sum_{a}\sum_{-\omega_{1}}\sum_{b}\sum_{\omega}e^{i\left(  \omega
_{1}-\omega \right)  t}\hat{A}_{a}(-\omega_{1})\hat{A}_{b}(\omega)\rho
_{I}^{\mathrm{S}}(t)\int_{0}^{t}dse^{i\omega s}tr_{\mathrm{B}}\left(  \hat
{B}_{a}^{\dag}(t)\hat{B}_{b}(t-s)\rho_{I}^{\mathrm{B}}\right) \nonumber \\
&  +\sum_{a}\sum_{\omega_{1}}\sum_{b}\sum_{\omega}e^{-i\left(  \omega
_{1}-\omega \right)  t}\hat{A}_{a}(\omega_{1})\rho_{I}^{\mathrm{S}}(t)\hat
{A}_{b}^{\dag}(\omega)\int_{0}^{t}dse^{-i\omega s}tr_{\text{\textrm{B}}%
}\left(  \rho_{I}^{\mathrm{B}}\hat{B}_{b}^{\dag}(t-s)\hat{B}_{a}(t)\right)
\nonumber \\
&  -\sum_{b}\sum_{\omega}\sum_{a}\sum_{-\omega_{1}}e^{-i\left(  \omega
_{1}-\omega \right)  t}\rho_{I}^{\mathrm{S}}(t)\hat{A}_{b}^{\dag}(\omega
)\hat{A}_{a}^{\dag}(-\omega_{1})\int_{0}^{t}dse^{-i\omega s}tr_{\mathrm{B}%
}\left(  \rho_{I}^{\mathrm{B}}\hat{B}_{b}^{\dag}(t-s)\hat{B}_{a}(t)\right)  .
\end{align}
We employ the reservoir correlation functions of the environment \textrm{B}
\begin{align*}
\Gamma_{ab}\left(  \omega \right)   &  =\int_{0}^{t}dse^{i\omega s}%
tr_{\mathrm{B}}\left(  \hat{B}_{a}^{\dag}(t)\hat{B}_{b}(t-s)\rho
_{I}^{\mathrm{B}}\right)  ,\\
\Gamma_{ba}^{\ast}\left(  \omega \right)   &  =\int_{0}^{t}dse^{-i\omega
s}tr_{\mathrm{B}}\left(  \rho_{I}^{\mathrm{B}}\hat{B}_{b}^{\dag}(t-s)\hat
{B}_{a}(t)\right)
\end{align*}
to simply the Eq.~\eqref{eq14} equation. Eventually, $\frac{d}{dt}\rho
_{I}^{\mathrm{S}}(t)$ of the subsystem \textrm{S} in the interaction picture
is%
\begin{align}
\frac{d}{dt}\rho_{I}^{\mathrm{S}}(t)  &  =\sum_{a,b}\sum_{\omega}\sum
_{\omega_{1}}e^{i\left(  \omega_{1}-\omega \right)  t}\Gamma_{ab}\left(
\omega \right)  \left(  \hat{A}_{b}(\omega)\rho_{I}^{\mathrm{S}}(t)A_{a}^{\dag
}(\omega_{1})-\hat{A}_{a}(-\omega_{1})\hat{A}_{b}(\omega)\rho_{I}^{\mathrm{S}%
}(t)\right) \nonumber \\
&  +\sum_{a,b}\sum_{\omega}\sum_{\omega_{1}}e^{-i\left(  \omega_{1}%
-\omega \right)  t}\Gamma_{ba}^{\ast}\left(  \omega \right)  \left(  \hat{A}%
_{a}(\omega_{1})\rho_{I}^{\mathrm{S}}(t)\hat{A}_{b}^{\dag}(\omega)-\rho
_{I}^{\mathrm{S}}(t)\hat{A}_{b}^{\dag}(\omega)\hat{A}_{a}^{\dag}(-\omega
_{1})\right)  .
\end{align}
We use the rotating wave approximation to average out the high frequency part
of quantum transition processes and ignore the case of $\omega \neq \omega_{1}$,
then get%
\begin{equation}
\frac{d}{dt}\rho_{I}^{\mathrm{S}}(t)=\sum_{a,b}\sum_{\omega}[\Gamma
_{ab}\left(  \omega \right)  (\hat{A}_{b}(\omega)\rho_{I}^{\mathrm{S}}%
(t)\hat{A}_{a}^{\dag}(\omega)-\hat{A}_{a}(-\omega)\hat{A}_{b}(\omega)\rho
_{I}^{\mathrm{S}}(t))+h.c.].
\end{equation}
Because the imaginary part of $\Gamma_{ab}$ only provides a shift of energy
levels and does not affect the NHTS that we will focus on, we ignore them in
the calculation below and get
\begin{equation}
\Gamma_{ab}=\frac{1}{2}\gamma_{ab},
\end{equation}
where $\gamma_{ab}=\int_{-\infty}^{\infty}dte^{i\omega t}\left \langle \hat
{B}_{a}^{\dag}(t)\hat{B}_{b}(0)\right \rangle =\int_{-\infty}^{\infty
}dte^{i\omega t}tr_{\mathrm{B}}\left(  \hat{B}_{a}^{\dag}(t)\hat{B}%
_{b}(t-s)\rho_{I}^{\mathrm{B}}\right)  $ is the real part of $\Gamma_{ab}$.
Using the Kubo-Martin-Schwinger (KMS) condition $\left \langle \hat{B}%
_{a}^{\dag}(t)\hat{B}_{b}(0)\right \rangle =\left \langle \hat{B}_{b}(0)\hat
{B}_{a}^{\dag}(t+i\frac{1}{T})\right \rangle $, we derive the temperature
dependent behavior of $\gamma_{ab},$ i.e.,
\begin{equation}
\gamma_{ab}(-\omega)=e^{-\omega/k_{B}T}\gamma_{ba}(\omega).
\end{equation}

For the effective NH Hamiltonian $\hat{H}_{\mathrm{NH}}$, we have $\omega
=\pm \omega_{0}$, $0$ with $\omega_{0}=|E_{+}-E_{-}|=2\sqrt{h^{2}-\gamma^{2}}$.
Thus, the GKSL equation becomes
\begin{align}
\frac{d}{dt}\rho_{I}^{\mathrm{S}}(t)  &  =\sum_{\omega}\Gamma_{11}\left(
\omega \right)  [\hat{A}_{1}(\omega)\rho_{I}^{\mathrm{S}}(t)\hat{A}_{1}^{\dag
}(\omega)-\hat{A}_{1}(-\omega)\hat{A}_{1}(\omega)\rho_{I}^{\mathrm{S}%
}(t)]\nonumber \\
&  +\sum_{\omega}\Gamma_{11}^{\ast}\left(  \omega \right)  [\hat{A}_{1}%
(\omega)\rho_{I}^{\mathrm{S}}(t)\hat{A}_{1}^{\dag}(\omega)-\rho_{I}%
^{\mathrm{S}}(t)\hat{A}_{1}^{\dag}(\omega)\hat{A}_{1}^{\dag}(-\omega
)]\nonumber \\
&  +\sum_{\omega}\Gamma_{22}\left(  \omega \right)  [\hat{A}_{2}(\omega
)\rho_{I}^{\mathrm{S}}(t)\hat{A}_{2}^{\dag}(\omega)-\hat{A}_{2}(-\omega
)\hat{A}_{2}(\omega)\rho_{I}^{\mathrm{S}}(t)]\nonumber \\
&  +\sum_{\omega}\Gamma_{22}^{\ast}\left(  \omega \right)  [\hat{A}_{2}%
(\omega)\rho_{I}^{\mathrm{S}}(t)\hat{A}_{2}^{\dag}(\omega)-\rho_{I}%
^{\mathrm{S}}(t)\hat{A}_{2}^{\dag}(\omega)\hat{A}_{2}^{\dag}(-\omega
)]\nonumber \\
&  +\sum_{\omega}\Gamma_{12}\left(  \omega \right)  [\hat{A}_{2}(\omega
)\rho_{I}^{\mathrm{S}}(t)\hat{A}_{1}^{\dag}(\omega)-\hat{A}_{1}(-\omega
)\hat{A}_{2}(\omega)\rho_{I}^{\mathrm{S}}(t)]\nonumber \\
&  +\sum_{\omega}\Gamma_{21}^{\ast}\left(  \omega \right)  [\hat{A}_{1}%
(\omega)\rho_{I}^{\mathrm{S}}(t)\hat{A}_{2}^{\dag}(\omega)-\rho_{I}%
^{\mathrm{S}}(t)\hat{A}_{2}^{\dag}(\omega)\hat{A}_{1}^{\dag}(-\omega
)]\nonumber \\
&  +\sum_{\omega}\Gamma_{21}\left(  \omega \right)  [\hat{A}_{1}(\omega
)\rho_{I}^{\mathrm{S}}(t)\hat{A}_{2}^{\dag}(\omega)-\hat{A}_{2}(-\omega
)\hat{A}_{1}(\omega)\rho_{I}^{\mathrm{S}}(t)]\nonumber \\
&  +\sum_{\omega}\Gamma_{12}^{\ast}\left(  \omega \right)  [\hat{A}_{2}%
(\omega)\rho_{I}^{\mathrm{S}}(t)\hat{A}_{1}^{\dag}(\omega)-\rho_{I}%
^{\mathrm{S}}(t)\hat{A}_{1}^{\dag}(\omega)\hat{A}_{2}^{\dag}(-\omega)].
\end{align}
We use $%
\begin{pmatrix}
A & B\\
C & D
\end{pmatrix}
$ to denote $\rho_{I}^{\mathrm{S}}(t).$ In combination with $\Gamma_{ab}%
=\frac{1}{2}\gamma_{ab}$ and $\hat{A}_{a,b}(\omega)$, we have the first term%
\begin{align}
&  \sum_{\omega}\Gamma_{11}\left(  \omega \right)  [\hat{A}_{1}(\omega)\rho
_{I}^{\mathrm{S}}(t)\hat{A}_{1}^{\dag}(\omega)-\hat{A}_{1}(-\omega)\hat{A}%
_{1}(\omega)\rho_{I}^{\mathrm{S}}(t)]\nonumber \\
&  =\Gamma_{11}\left(  \omega_{0}\right)  [\hat{A}_{1}(\omega_{0})\rho
_{I}^{\mathrm{S}}(t)\hat{A}_{1}^{\dag}(\omega_{0})-\hat{A}_{1}(-\omega
_{0})\hat{A}_{1}(\omega_{0})\rho_{I}^{\mathrm{S}}(t)]\nonumber \\
&  +\Gamma_{11}\left(  -\omega_{0}\right)  [\hat{A}_{1}(-\omega_{0})\rho
_{I}^{\mathrm{S}}(t)\hat{A}_{1}^{\dag}(-\omega_{0})-\hat{A}_{1}(\omega
_{0})\hat{A}_{1}(-\omega_{0})\rho_{I}^{\mathrm{S}}(t)]\nonumber \\
&  +\Gamma_{11}\left(  0\right)  [\hat{A}_{1}(0)\rho_{I}^{\mathrm{S}}%
(t)\hat{A}_{1}^{\dag}(0)-\hat{A}_{1}(0)\hat{A}_{1}(0)\rho_{I}^{\mathrm{S}%
}(t)]\nonumber \\
&  =\frac{1}{32}e^{\omega_{0}/k_{B}T}\gamma_{11}\left(  -\omega_{0}\right)
\gamma_{1}^{2}\left(
\begin{array}
[c]{cc}%
-A+xB-xC+x^{2}D & -Ay-3B-C-3xD\\
-3yA-B-3C-xD & y^{2}A-yB+yC-D
\end{array}
\right) \nonumber \\
&  +\frac{1}{32}\gamma_{11}\left(  -\omega_{0}\right)  \gamma_{1}^{2}\left(
\begin{array}
[c]{cc}%
-A-xB+xC+x^{2}D & Ay-3B-C+3xD\\
3Ay-B-3C+xD & y^{2}A+yB-yC-D
\end{array}
\right)  .
\end{align}
The second term is%
\begin{align}
&  \sum_{\omega}\Gamma_{11}^{\ast}\left(  \omega \right)  [\hat{A}_{1}%
(\omega)\rho_{I}^{\mathrm{S}}(t)\hat{A}_{1}^{\dag}(\omega)-\rho_{I}%
^{\mathrm{S}}(t)\hat{A}_{1}^{\dag}(\omega)\hat{A}_{1}^{\dag}(-\omega
)]\nonumber \\
&  =\frac{1}{32}e^{\omega_{0}/k_{B}T}\gamma_{11}\left(  -\omega_{0}\right)
\gamma_{1}^{2}\left(
\begin{array}
[c]{cc}%
-A-xB+xC+x^{2}D & -3Ay-3B-C-xD\\
-yA-B-3C-3yD & y^{2}A+yB-yC-D
\end{array}
\right) \nonumber \\
&  +\frac{1}{32}\gamma_{11}\left(  -\omega_{0}\right)  \gamma_{1}^{2}\left(
\begin{array}
[c]{cc}%
-A+xB-xC+x^{2}D & 3Ab-3B-C+xD\\
yA-B-3C+3xD & y^{2}A-yB+yC-D
\end{array}
\right)  .
\end{align}
The third term is%
\begin{align}
&  \sum_{\omega}\Gamma_{22}\left(  \omega \right)  [A_{2}(\omega)\rho
_{I}^{\mathrm{S}}(t)A_{2}^{\dag}(\omega)-A_{2}(-\omega)A_{2}(\omega)\rho
_{I}^{\mathrm{S}}(t)]\nonumber \\
&  =\frac{1}{32}e^{\omega_{0}/k_{B}T}\gamma_{22}\left(  -\omega_{0}\right)
\gamma_{2}^{2}\left(
\begin{array}
[c]{cc}%
-A+xB-xC+x^{2}D & -Ay-3B-C-3xD\\
-3yA-B-3C-xD & y^{2}A-yB+yC-D
\end{array}
\right) \nonumber \\
&  +\frac{1}{32}\gamma_{22}\left(  -\omega_{0}\right)  \gamma_{2}^{2}\left(
\begin{array}
[c]{cc}%
-A-xB+xC+x^{2}D & Ay-3B-C+3xD\\
3Ay-B-3C+xD & y^{2}A+yB-yC-D
\end{array}
\right)  .
\end{align}
The fourth term is%
\begin{align}
&  \sum_{\omega}\Gamma_{22}^{\ast}\left(  \omega \right)  [\hat{A}_{2}%
(\omega)\rho_{I}^{\mathrm{S}}(t)\hat{A}_{2}^{\dag}(\omega)-\rho_{I}%
^{\mathrm{S}}(t)\hat{A}_{2}^{\dag}(\omega)\hat{A}_{2}^{\dag}(-\omega
)]\nonumber \\
&  =\frac{1}{32}e^{\omega_{0}/k_{B}T}\gamma_{22}\left(  -\omega_{0}\right)
\gamma_{2}^{2}\left(
\begin{array}
[c]{cc}%
-A-xB+xC+x^{2}D & -3yA-3B-C-xD\\
-Ay-B-3C-3xD & y^{2}A+yB-yC-D
\end{array}
\right) \nonumber \\
&  +\frac{1}{32}\gamma_{22}\left(  -\omega_{0}\right)  \gamma_{2}^{2}\left(
\begin{array}
[c]{cc}%
-A+xB-xC+x^{2}D & 3Ay-3B-C+xD\\
Ay-B-3C+3xD & y^{2}A-yB+yC-D
\end{array}
\right)  .
\end{align}
The fifth term is%
\begin{align}
&  \sum_{\omega}\Gamma_{12}\left(  \omega \right)  [\hat{A}_{2}(\omega)\rho
_{I}^{\mathrm{S}}(t)\hat{A}_{1}^{\dag}(\omega)-\hat{A}_{1}(-\omega)\hat{A}%
_{2}(\omega)\rho_{I}^{\mathrm{S}}(t)]\nonumber \\
&  =-\frac{1}{32}e^{\omega_{0}/k_{B}T}\gamma_{21}\left(  -\omega_{0}\right)
\gamma_{1}\gamma_{2}\left(
\begin{array}
[c]{cc}%
-A+xB-xC+x^{2}D & -Ay-3B-C-3xD\\
-3yA-B-3C-xD & y^{2}A-yB+yC-D
\end{array}
\right) \nonumber \\
&  -\frac{1}{32}\gamma_{12}\left(  -\omega_{0}\right)  \gamma_{1}\gamma
_{2}\left(
\begin{array}
[c]{cc}%
-A-xB+xC+x^{2}D & Ay-3B-C+3xD\\
3Ay-B-3C+xD & y^{2}A+yB-yC-D
\end{array}
\right)  .
\end{align}
The sixth term is%
\begin{align}
&  \sum_{\omega}\Gamma_{21}^{\ast}\left(  \omega \right)  [\hat{A}_{1}%
(\omega)\rho_{I}^{\mathrm{S}}(t)\hat{A}_{2}^{\dag}(\omega)-\rho_{I}%
^{\mathrm{S}}(t)\hat{A}_{2}^{\dag}(\omega)\hat{A}_{1}^{\dag}(-\omega
)]\nonumber \\
&  =-\frac{1}{32}e^{\omega_{0}/k_{B}T}\gamma_{21}\left(  -\omega_{0}\right)
\gamma_{1}\gamma_{2}\left(
\begin{array}
[c]{cc}%
-A-xB+xC+x^{2}D & -3yA-3B-C-xD\\
-Ay-B-3C-3xD & y^{2}A+yB-yC-D
\end{array}
\right) \nonumber \\
&  -\frac{1}{32}\gamma_{12}\left(  -\omega_{0}\right)  \gamma_{1}\gamma
_{2}\left(
\begin{array}
[c]{cc}%
-A+xB-xC+x^{2}D & 3Ay-3B-C+xD\\
Ay-B-3C+3xD & y^{2}A-yB+yC-D
\end{array}
\right)  .
\end{align}
The seventh term is%
\begin{align}
&  \sum_{\omega}\Gamma_{21}\left(  \omega \right)  [\hat{A}_{1}(\omega)\rho
_{I}^{\mathrm{S}}(t)\hat{A}_{2}^{\dag}(\omega)-\hat{A}_{2}(-\omega)\hat{A}%
_{1}(\omega)\rho_{I}^{\mathrm{S}}(t)]\nonumber \\
&  =-\frac{1}{32}e^{\omega_{0}/k_{B}T}\gamma_{12}\left(  -\omega_{0}\right)
\gamma_{1}\gamma_{2}\left(
\begin{array}
[c]{cc}%
-A+xB-xC+x^{2}D & -Ay-3B-C-3xD\\
-3yA-B-3C-xD & y^{2}A-yB+yC-D
\end{array}
\right) \nonumber \\
&  -\frac{1}{32}\gamma_{21}\left(  -\omega_{0}\right)  \gamma_{1}\gamma
_{2}\left(
\begin{array}
[c]{cc}%
-A-xB+xC+x^{2}D & Ay-3B-C+3xD\\
3Ay-B-3C+xD & y^{2}A+yB-yC-D
\end{array}
\right)  .
\end{align}
The eighth term is%
\begin{align}
&  \sum_{\omega}\Gamma_{12}^{\ast}\left(  \omega \right)  [\hat{A}_{2}%
(\omega)\rho_{I}^{\mathrm{S}}(t)\hat{A}_{1}^{\dag}(\omega)-\rho_{I}%
^{\mathrm{S}}(t)\hat{A}_{1}^{\dag}(\omega)\hat{A}_{2}^{\dag}(-\omega
)]\nonumber \\
&  =-\frac{1}{32}e^{\omega_{0}/k_{B}T}\gamma_{12}\left(  -\omega_{0}\right)
\gamma_{1}\gamma_{2}\left(
\begin{array}
[c]{cc}%
-A-xB+xC+x^{2}D & -3yA-3B-C-xD\\
-Ay-B-3C-3xD & y^{2}A+yB-yC-D
\end{array}
\right) \nonumber \\
&  -\frac{1}{32}\gamma_{21}\left(  -\omega_{0}\right)  \gamma_{1}\gamma
_{2}\left(
\begin{array}
[c]{cc}%
-A+xB-xC+x^{2}D & 3Ay-3B-C+xD\\
Ay-B-3C+3xD & y^{2}A-yB+yC-D
\end{array}
\right)  .
\end{align}
Combining them together, we get%
\begin{align}
\frac{d}{dt}\rho_{I}^{\mathrm{S}}(t)  &  =\gamma_{0}e^{\omega_{0}/k_{B}%
T}\left(
\begin{array}
[c]{cc}%
-A+x^{2}D & -2Ay-3B-C-2xD\\
-2yA-B-3C-2xD & y^{2}A-D
\end{array}
\right) \nonumber \\
&  +\gamma_{0}\left(
\begin{array}
[c]{cc}%
-A+x^{2}D & 2yA-3B-C+2xD\\
2yA-B-3C+2xD & y^{2}A-D
\end{array}
\right)  ,
\end{align}
where
\begin{equation}
\gamma_{0}=\frac{1}{16}[\gamma_{11}\left(  -\omega_{0}\right)  \gamma_{1}%
^{2}+\gamma_{22}\left(  -\omega_{0}\right)  \gamma_{2}^{2}-\left(  \gamma
_{12}\left(  -\omega_{0}\right)  +\gamma_{21}\left(  -\omega_{0}\right)
\right)  \gamma_{1}\gamma_{2}].
\end{equation}
According to the quantum master equation in the GKSL form $\frac{d\rho
_{I}^{\mathrm{S}}(t)}{dt}\equiv \mathcal{L}\rho_{I}^{\mathrm{S}}(t)$, we derive%

\begin{equation}
\frac{dA}{dt}=-\gamma_{0}\left(  e^{\omega_{0}/k_{B}T}+1\right)  A+\gamma
_{0}\left(  e^{\omega_{0}/k_{B}T}+1\right)  D,
\end{equation}%
\begin{align}
\frac{dB}{dt}  &  =-2\gamma_{0}y\left(  e^{\omega_{0}/k_{B}T}-1\right)
A-3\gamma_{0}\left(  e^{\omega_{0}/k_{B}T}+1\right)  B\nonumber \\
&  -\gamma_{0}\left(  e^{\omega_{0}/k_{B}T}+1\right)  C-2\gamma_{0}x\left(
e^{\omega_{0}/k_{B}T}-1\right)  D,
\end{align}%
\begin{align}
\frac{dC}{dt}  &  =-2\gamma_{0}y\left(  e^{\omega_{0}/k_{B}T}-1\right)
A-\gamma_{0}\left(  e^{\omega_{0}/k_{B}T}+1\right)  B\nonumber \\
&  -3\gamma_{0}\left(  e^{\omega_{0}/k_{B}T}+1\right)  C-2\gamma_{0}x\left(
e^{\omega_{0}/k_{B}T}-1\right)  D,
\end{align}%
\begin{equation}
\frac{dD}{dt}=\gamma_{0}\left(  e^{\omega_{0}/k_{B}T}+1\right)  b^{2}%
A-\gamma_{0}\left(  e^{\omega_{0}/k_{B}T}+1\right)  D.
\end{equation}
Eventually, we have
\begin{equation}
\mathcal{L=}\left(
\begin{array}
[c]{cccc}%
-\gamma_{0}\left(  1+e^{\omega_{0}/k_{B}T}\right)  & 0 & 0 & \gamma_{0}\left(
1+e^{\omega_{0}/k_{B}T}\right)  \frac{h+\gamma}{h-\gamma}\\
-2\gamma_{0}\left(  e^{\omega_{0}/k_{B}T}-1\right)  \sqrt{\frac{h-\gamma
}{h+\gamma}} & -3\gamma_{0}\left(  1+e^{\omega_{0}/k_{B}T}\right)  &
-\gamma_{0}\left(  1+e^{\omega_{0}/k_{B}T}\right)  & -2\gamma_{0}\left(
e^{\omega_{0}/k_{B}T}-1\right)  \sqrt{\frac{h+\gamma}{h-\gamma}}\\
-2\gamma_{0}\left(  e^{\omega_{0}/k_{B}T}-1\right)  \sqrt{\frac{h-\gamma
}{h+\gamma}} & -\gamma_{0}\left(  1+e^{\omega_{0}/k_{B}T}\right)  &
-3\gamma_{0}\left(  1+e^{\omega_{0}/k_{B}T}\right)  & -2\gamma_{0}\left(
e^{\omega_{0}/k_{B}T}-1\right)  \sqrt{\frac{h+\gamma}{h-\gamma}}\\
\gamma_{0}\left(  1+e^{\omega_{0}/k_{B}T}\right)  \frac{h-\gamma}{h+\gamma} &
0 & 0 & -\gamma_{0}\left(  1+e^{\omega_{0}/k_{B}T}\right)
\end{array}
\right)  .
\end{equation}

Thirdly, we derive the reduced density matrix $\rho_{NHTS}^{\mathrm{S}}$ for
the NHTS according to the Liouville super-operator $\mathcal{L}$.

The eigenvalues of $\mathcal{L}$ are $-4\gamma_{0}(1+e^{\omega_{0}/T})$,
$-2\gamma_{0}(1+e^{\omega_{0}/T})$ (double degeneracy), and $0$ in the
interaction picture. We have $\gamma_{0}>0$ because $\gamma_{0}$ and $\gamma$
are independent. For the Hermitian system $\gamma=0$, which the postselection
measurement does not affect system evolution, the maximum eigenvalue
of\textbf{ }$\mathcal{L}$\textbf{ }must be\textbf{ }$0$.\textbf{ }The first
three will dissipate, the rest $0$ will form a NHTS that is also a steady
state. Thus, the density matrix $\rho_{NHTS}^{\mathrm{S}}$ for a NHTS
corresponding to the eigenvalue of the Liouville super-operator $\mathcal{L}$
with maximum real part is obtained as%
\begin{equation}
\rho_{NHTS}^{\mathrm{S}}=\frac{h-\gamma}{2h}%
\begin{pmatrix}
\frac{h+\gamma}{h-\gamma} & \sqrt{\frac{h+\gamma}{h-\gamma}}\frac
{1-e^{\omega_{0}\beta_{T}}}{1+e^{\omega_{0}\beta_{T}}}\\
\sqrt{\frac{h+\gamma}{h-\gamma}}\frac{1-e^{\omega_{0}\beta_{T}}}%
{1+e^{\omega_{0}\beta_{T}}} & 1
\end{pmatrix}
.
\end{equation}

\section{The generalized Baker--Campbell--Hausdorff formulation}

\emph{Generalized Baker--Campbell--Hausdorff formulation}\textbf{:}%
\textit{\ }For the case of $\sigma_{1}^{2}=1$ and $\sigma_{2}^{2}=1$, we have
the following equation,%
\begin{equation}
\rho=e^{\vec{r}_{1}\cdot \vec{\sigma}}\cdot e^{\vec{r}_{2}\cdot \vec{\sigma}%
}\cdot e^{\vec{r}_{1}\cdot \vec{\sigma}}=\alpha+\left \vert \vec{r}%
_{\mathrm{eff}}\right \vert \cdot \sigma_{\mathrm{eff}}=e^{c+\vec{r}\cdot
\vec{\sigma}},
\end{equation}
where
\begin{equation}
c=\frac{1}{2}\ln(\alpha^{2}-\left \vert \vec{r}_{\mathrm{eff}}\right \vert ^{2})
\end{equation}
and
\begin{equation}
\left \vert \vec{r}\right \vert =\tanh^{-1}(\frac{\left \vert \vec{r}%
_{\mathrm{eff}}\right \vert }{\alpha})
\end{equation}
with
\[
\alpha=\cosh \left(  2r_{1}\right)  \cdot \cosh r_{2}+(\vec{n}_{1}\cdot \vec
{n}_{2})\sinh \left(  2r_{1}\right)  \cdot \sinh r_{2},
\]%
\[
\left \vert \vec{r}_{\mathrm{eff}}\right \vert =\sqrt{\left[  \sinh \left(
2r_{1}\right)  \cdot \cosh r_{2}+2\left(  \vec{n}_{1}\cdot \vec{n}_{2}\right)
\sinh^{2}r_{1}\cdot \sinh r_{2}\right]  ^{2}+(\sinh r_{2})^{2}},
\]%
\[
\sigma_{\mathrm{eff}}=\left \vert \vec{r}_{\mathrm{eff}}\right \vert ^{-1}%
\cdot(\sigma_{1}[\sinh \left(  2r_{1}\right)  \cdot \cosh r_{2}+2\left(  \vec
{n}_{1}\cdot \vec{n}_{2}\right)  \sinh^{2}r_{1}\cdot \sinh r_{2}]+\sigma
_{2}\sinh r_{2}),
\]
$r_{1}=\left \vert \vec{r}_{1}\right \vert $ and $r_{2}=\left \vert \vec{r}%
_{2}\right \vert $.

We then prove the above generalized Baker--Campbell--Hausdorff (BCH) formulation.

Firstly, for $\sigma_{n}^{2}=1$, we have
\begin{equation}
e^{\vec{r}\cdot \vec{\sigma}}=\cosh r+\sigma_{n}\sinh r,
\end{equation}
where $\left \vert \vec{r}\right \vert =r$ with $\cosh^{2}r-\sinh^{2}r=1$ and
$\sigma_{n}=\frac{\vec{r}\cdot \vec{\sigma}}{r}$ . On the other hand, for the
case of $X=\vec{d}+\vec{e}\cdot \vec{\sigma}$ with $e=\left \vert \vec
{e}\right \vert ,$ we have $X=e^{c+\vec{r}\cdot \vec{\sigma}}$ where
\begin{equation}
e^{c}\cosh r=d,\text{ }e^{c}\sinh r=e.
\end{equation}
Then, we have $r=\tanh^{-1}(\frac{e}{d})$ and $c=\frac{1}{2}\ln(d^{2}-e^{2}).$

Next, we calculate $\rho=e^{\vec{r}_{1}\cdot \vec{\sigma}}\cdot e^{\vec{r}%
_{2}\cdot \vec{\sigma}}\cdot e^{\vec{r}_{1}\cdot \vec{\sigma}}$ and get%
\begin{align}
\rho &  =e^{\vec{r}_{1}\cdot \vec{\sigma}}\cdot e^{\vec{r}_{2}\cdot \vec{\sigma
}}\cdot e^{\vec{r}_{1}\cdot \vec{\sigma}}\nonumber \\
&  =\left(  \cosh r_{1}+\sigma_{1}\sinh r_{1}\right)  \nonumber \\
&  \cdot \left(  \cosh r_{2}+\sigma_{2}\sinh r_{2}\right)  \nonumber \\
&  \cdot \left(  \cosh r_{1}+\sigma_{1}\sinh r_{1}\right)  ,
\end{align}
where $r_{1}=\left \vert \vec{r}_{1}\right \vert $ and $r_{2}=\left \vert \vec
{r}_{2}\right \vert $. The result is obtained as
\begin{align}
\rho &  =(\cosh r_{1}\cdot \cosh r_{2}+\sigma_{1}\sinh r_{1}\cdot \cosh
r_{2}\nonumber \\
&  +\sigma_{2}\cosh r_{1}\cdot \sinh r_{2}+\sigma_{1}\sigma_{2}\sinh r_{1}%
\cdot \sinh r_{2})\nonumber \\
&  \cdot \left(  \cosh r_{1}+\sigma_{1}\sinh r_{1}\right)  \nonumber \\
&  =\cosh \left(  2r_{1}\right)  \cdot \cosh r_{2}+\left(  \vec{n}_{1}\cdot
\vec{n}_{2}\right)  \sinh \left(  2r_{1}\right)  \cdot \sinh r_{2}\nonumber \\
&  +\sigma_{1}[\sinh \left(  2r_{1}\right)  \cdot \cosh r_{2}+2\left(  \vec
{n}_{1}\cdot \vec{n}_{2}\right)  \sinh^{2}r_{1}\cdot \sinh r_{2}]\nonumber \\
&  +\sigma_{2}\sinh r_{2},
\end{align}
where $\vec{n}_{1}=\frac{\vec{r}_{1}}{\left \vert \vec{r}_{1}\right \vert }$ and
$\vec{n}_{2}=\frac{\vec{r}_{2}}{\left \vert \vec{r}_{2}\right \vert }$. Here, we
have used $\{ \sigma_{1},\sigma_{2}\}=2(\vec{n}_{1}\cdot \vec{n}_{2}).$

Thirdly, we show the final result:
\begin{align}
\rho &  =\alpha+\left \vert \vec{r}_{\mathrm{eff}}\right \vert \cdot
\sigma_{\mathrm{eff}}\nonumber \\
&  =e^{c+\vec{r}\cdot \vec{\sigma}},
\end{align}
where%
\begin{align*}
c &  =\frac{1}{2}\ln(\alpha^{2}-\left \vert \vec{r}_{\mathrm{eff}}\right \vert
^{2}),\\
r &  =\tanh^{-1}(\frac{\left \vert \vec{r}_{\mathrm{eff}}\right \vert }{\alpha})
\end{align*}
with $\alpha=\cosh2r_{1}\cdot \cosh r_{2}+\left(  \vec{n}_{1}\cdot \vec{n}%
_{2}\right)  \sinh2r_{1}\cdot \sinh r_{2}$,
\begin{equation}
\left \vert \vec{r}_{\mathrm{eff}}\right \vert =\sqrt{\left[  \sinh \left(
2r_{1}\right)  \cdot \cosh r_{2}+2\left(  \vec{n}_{1}\cdot \vec{n}_{2}\right)
\sinh^{2}r_{1}\cdot \sinh r_{2}\right]  ^{2}+(\sinh r_{2})^{2}}\nonumber
\end{equation}
and
\begin{equation}
\sigma_{\mathrm{eff}}=\left \vert \vec{r}_{\mathrm{eff}}\right \vert ^{-1}%
\cdot \left \{  \sigma_{1}[\sinh \left(  2r_{1}\right)  \cdot \cosh r_{2}+2\left(
\vec{n}_{1}\cdot \vec{n}_{2}\right)  \sinh^{2}r_{1}\cdot \sinh r_{2}]+\sigma
_{2}\sinh r_{2}\right \}  .\nonumber
\end{equation}
In general, according to $\alpha^{2}-\left \vert \vec{r}_{\mathrm{eff}%
}\right \vert ^{2}=1$, we have
\begin{equation}
c=\frac{1}{2}\ln(\alpha^{2}-\left \vert \vec{r}_{\mathrm{eff}}\right \vert
^{2})\equiv0.
\end{equation}
So%
\begin{equation}
\rho=e^{\frac{\cosh^{-1}\alpha}{\left \vert \vec{r}_{\mathrm{eff}}\right \vert
}\left(  \vec{\sigma}\cdot \vec{r}_{\mathrm{eff}}\right)  }=e^{\cosh^{-1}%
\alpha \cdot \sigma_{\mathrm{eff}}}.
\end{equation}

\section{The detailed calculations of the expected values of physical
operators $\sigma_{i}$}

In the case of real energy levels ($h>\gamma$), we have a steady state (or the
NHTS for this case) for the original NH system $\hat{H}_{\mathrm{NH}}$ at
finite temperature. By doing similarity transformation (ST) $\mathcal{\hat{S}%
}=e^{\beta_{\mathrm{NH}}\hat{H}^{\prime}},$ $\hat{H}_{\mathrm{NH}}$ can be
transformed into a Hermitian Hamiltonian
\begin{equation}
\hat{H}_{0}=\sqrt{h^{2}-\gamma^{2}}\sigma_{x},
\end{equation}
i.e.,
\begin{equation}
\hat{H}_{\mathrm{NH}}=\mathcal{\hat{S}}\hat{H}_{0}\mathcal{\hat{S}}^{-1}.
\end{equation}
Here, the real number $\beta_{\mathrm{NH}}=\frac{1}{2}\ln \frac{h+\gamma
}{h-\gamma}$ characterizes the strength of NH terms, and the Hermitian
operator $\hat{H}^{\prime}=\frac{1}{2}\sigma_{z}$ determines the form of the
NH terms.

According to the definition of $e^{-\beta_{T}\hat{H}_{\mathrm{L}}%
}=\mathcal{\hat{S}}(e^{-\beta_{T}\hat{H}_{0}})\mathcal{\hat{S}}^{\dagger},$ we
have
\begin{equation}
e^{-\beta_{T}\hat{H}_{\mathrm{L}}}=\sigma_{x}A+\sigma_{z}B+\mathrm{I}C
\end{equation}
where $\mathrm{I}=(%
\begin{array}
[c]{cc}%
1 & 0\\
0 & 1
\end{array}
),$
\begin{align}
A &  =-\sinh(\beta_{T}\sqrt{h^{2}-\gamma^{2}}),\nonumber \\
B &  =\sinh(\beta_{\mathrm{NH}})\cdot \cosh(\beta_{T}\sqrt{h^{2}-\gamma^{2}%
}),\nonumber \\
C &  =\cosh(\beta_{\mathrm{NH}})\cdot \cosh(\beta_{T}\sqrt{h^{2}-\gamma^{2}}).
\end{align}
From the generalized BCH formulation, the Liouvillian Hamiltonian is obtained
as%
\begin{equation}
\hat{H}_{\mathrm{L}}=-\frac{1}{\beta_{T}}\frac{\cosh^{-1}C}{\left \vert \vec
{r}_{\mathrm{eff}}\right \vert }(\vec{\sigma}\cdot \vec{r}_{\mathrm{eff}}),
\end{equation}
where $\vec{r}_{\mathrm{eff}}=(A,0,B)$, $\left \vert \vec{r}_{\mathrm{eff}%
}\right \vert =\sqrt{A^{2}+B^{2}}$, and $\vec{\sigma}=(\sigma_{x},\sigma
_{y},\sigma_{z}).$ Specifically,
\begin{align}
\hat{H}_{\mathrm{L}} &  =\frac{\cosh^{-1}\left[  \cosh \beta_{\mathrm{NH}}%
\cosh \left(  \beta_{T}\sqrt{h^{2}-\gamma^{2}}\right)  \right]  }{\beta
_{T}\sqrt{\left[  \sinh \left(  \beta_{T}\sqrt{h^{2}-\gamma^{2}}\right)
\right]  ^{2}+\left[  \sinh \beta_{\mathrm{NH}}\cosh \left(  \beta_{T}%
\sqrt{h^{2}-\gamma^{2}}\right)  \right]  ^{2}}}\nonumber \\
&  \cdot \left[  \sinh \left(  \beta_{T}\sqrt{h^{2}-\gamma^{2}}\right)
\cdot \sigma_{x}-\sinh \beta_{\mathrm{NH}}\cosh \left(  \beta_{T}\sqrt
{h^{2}-\gamma^{2}}\right)  \cdot \sigma_{z}\right]  .
\end{align}
After diagonalizing $\hat{H}_{\mathrm{L}}$, the Liouvillian energy levels
$E_{\pm}^{\mathrm{L}}$ are%
\begin{equation}
E_{\pm}^{\mathrm{L}}=\pm \frac{1}{\beta_{T}}\cosh^{-1}\left[  \cosh
\beta_{\mathrm{NH}}\cdot \cosh(\beta_{T}\sqrt{h^{2}-\gamma^{2}})\right]
\end{equation}
that are quite different from the energy levels (the eigenvalues for $\hat
{H}_{\mathrm{NH}}$) $E_{\pm}=\pm \sqrt{h^{2}-\gamma^{2}}$, as shown in Fig.1.
As a result, the Liouvillian energy gap is
\begin{equation}
\Delta=2\left \vert \varepsilon \right \vert =\frac{2}{\beta_{T}}\cosh
^{-1}\left[  \cosh \beta_{\mathrm{NH}}\cdot \cosh(\beta_{T}\sqrt{h^{2}%
-\gamma^{2}})\right]  .
\end{equation}
At the EP, $\Delta$ diverges, i.e., when $\beta_{\mathrm{NH}}\rightarrow
\infty,$ $\Delta \rightarrow \infty$. Remember the energy gap for $\hat
{H}_{\mathrm{NH}}$ at EP turns to zero, i.e., when $\beta_{\mathrm{NH}%
}\rightarrow \infty,$ $\omega_{0}=2\sqrt{h^{2}-\gamma^{2}}\rightarrow0$.
\begin{figure}[ptb]
\includegraphics[clip,width=0.9\textwidth]{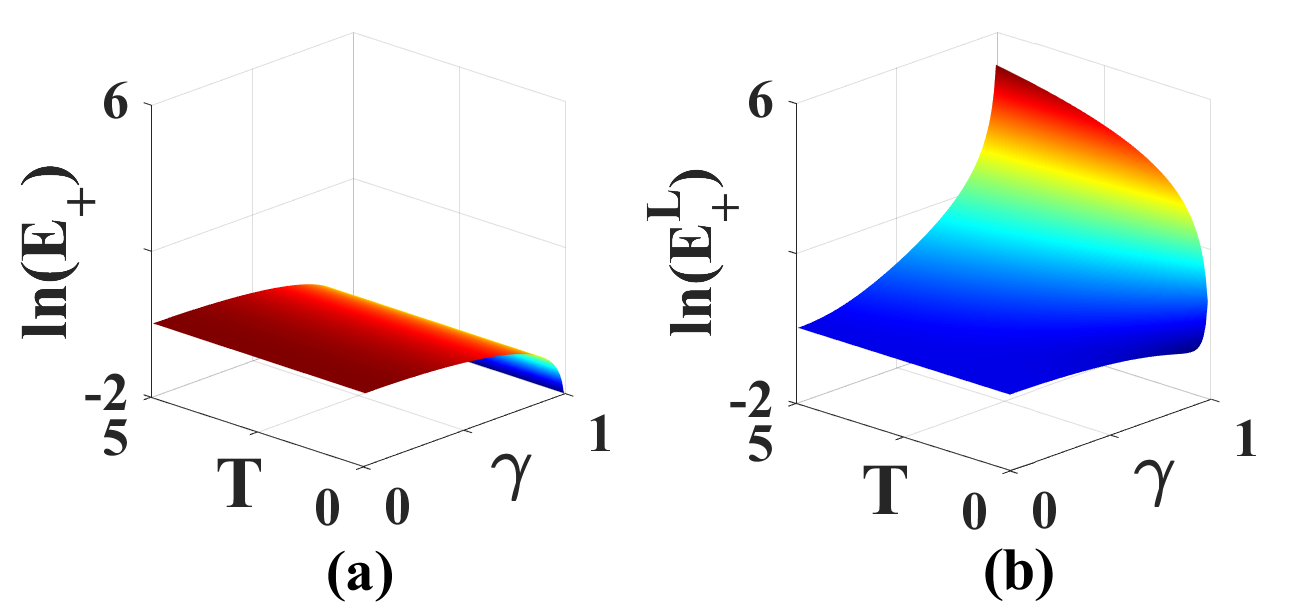}\caption{(a) The energy
level $E_{+}=\sqrt{h^{2}-\gamma^{2}}$ versus temperature $T$ and $\gamma$ for
non-Hermitian Hamiltonian $\hat{H}_{\mathrm{NH}}$. (b) The Liouvillian energy
level $E_{+}^{\mathrm{L}}=\frac{1}{\beta_{T}}\cosh^{-1}\left[  \cosh
\beta_{\mathrm{NH}}\cdot \cosh(\beta_{T}\sqrt{h^{2}-\gamma^{2}})\right]  $
versus temperature $T$ and $\gamma$ for the Liouvillian Hamiltonian $\hat
{H}_{\mathrm{L}}$.}%
\end{figure}

The partition function $Z_{NHTS}^{\mathrm{S}}$ is given by
\begin{align}
Z_{NHTS}^{\mathrm{S}}  &  =\mathrm{Tr}(\rho_{NHTS}^{\mathrm{S}})=\mathrm{Tr}%
(e^{-\beta_{T}\hat{H}_{\mathrm{L}}})\nonumber \\
&  =\mathrm{Tr}(\sigma_{x}A+\sigma_{z}B+\mathrm{I}C)=2C.
\end{align}
We then calculate the expected values of $\sigma_{i}$ ($i=x,y,z$), i.e.,
$n_{i}=\left \langle \sigma_{i}\right \rangle .$ The expected value of physical
operator $\sigma_{i}$ is defined as
\begin{align}
\left \langle \sigma_{i}\right \rangle  &  =\frac{1}{Z_{NHTS}^{\mathrm{S}}%
}\mathrm{Tr}(\sigma_{i}\cdot e^{-\beta_{T}\hat{H}_{\mathrm{L}}})\nonumber \\
&  =\frac{1}{Z_{NHTS}^{\mathrm{S}}}\mathrm{Tr}\left[  \sigma_{i}\cdot
(\sigma_{x}A+\sigma_{y}B+\mathrm{I}C)\right]  .
\end{align}
After straightforward calculations, we have
\begin{align}
\left \langle \sigma_{x}\right \rangle  &  =\frac{A}{C}=\frac{-\tanh \left(
\beta_{T}\sqrt{h^{2}-\gamma^{2}}\right)  }{\cosh \beta_{\mathrm{NH}}%
},\nonumber \\
\left \langle \sigma_{y}\right \rangle  &  =0,\nonumber \\
\left \langle \sigma_{z}\right \rangle  &  =\frac{B}{C}=\tanh \beta_{\mathrm{NH}%
}.
\end{align}

In the case of imaginary energy levels ($h<\gamma$), the NHTS $\rho
_{NHTS}^{\mathrm{S}}$ is%
\begin{align}
\rho_{NHTS}^{\mathrm{S}} &  =|\Psi_{+}^{\mathrm{R}}\rangle \langle \Psi
_{+}^{\mathrm{R}}|\nonumber \\
&  =\frac{1}{2}\binom{1}{-\sqrt{\frac{h-\gamma}{h+\gamma}}}\left(
\begin{array}
[c]{cc}%
1 & \left(  -\sqrt{\frac{h-\gamma}{h+\gamma}}\right)  ^{\ast}%
\end{array}
\right)  \nonumber \\
&  =\frac{1}{2}\left(
\begin{array}
[c]{cc}%
1 & -\sqrt{\frac{h-\gamma}{h+\gamma}}\\
-\sqrt{\frac{h-\gamma}{h+\gamma}} & \frac{\gamma-h}{h+\gamma}%
\end{array}
\right)  .
\end{align}
Then, its partition function $Z_{NHTS}^{\mathrm{S}}$ is
\begin{equation}
Z_{NHTS}^{\mathrm{S}}=\mathrm{Tr}(\rho_{NHTS}^{\mathrm{S}})=\frac{1}{2}%
+\frac{1}{2}\frac{\gamma-h}{h+\gamma}=\frac{\gamma}{h+\gamma}.
\end{equation}
Similarly, we use $\left \langle \sigma_{i}\right \rangle =\frac{1}%
{Z_{NHTS}^{\mathrm{S}}}\mathrm{Tr}(\sigma_{i}\cdot \rho_{NHTS}^{\mathrm{S}})$
and can see that%

\begin{align}
\left \langle \sigma_{x}\right \rangle  &  =0,\nonumber \\
\left \langle \sigma_{y}\right \rangle  &  =-\frac{\sqrt{\gamma^{2}-h^{2}}%
}{\gamma},\nonumber \\
\left \langle \sigma_{z}\right \rangle  &  =\frac{h}{\gamma}.
\end{align}

\section{The detailed calculations for thermodynamic phase transition at the
exceptional point}

According to the results in main content, the average spin operator $\vec
{n}=(n_{x},n_{y},n_{z})=(\left \langle \sigma_{x}\right \rangle ,\left \langle
\sigma_{y}\right \rangle ,\left \langle \sigma_{z}\right \rangle )$ is obtained
as%
\begin{align}
\vec{n}  &  =(\frac{A}{C},0,\frac{B}{C})\nonumber \\
&  =(\frac{-\tanh \left(  \beta_{T}\sqrt{h^{2}-\gamma^{2}}\right)  }{\cosh
\beta_{\mathrm{NH}}},0,\tanh \beta_{\mathrm{NH}})
\end{align}
for $h>\gamma$ and
\begin{equation}
\vec{n}=(0,-\sqrt{1-(\frac{h}{\gamma})^{2}},\frac{h}{\gamma})
\end{equation}
for $h<\gamma.$

We can see that the average spin operator $\left \langle \sigma_{i}%
\right \rangle $ is always continuous crossing over the exceptional point (EP).
So we calculate the derivatives of spin average values $\frac{\partial \vec{n}%
}{\partial \gamma}$,%
\begin{align}
\frac{\partial n_{x}}{\partial \gamma}  &  =\left \{  \frac{1}{h}\tanh \left(
\beta_{T}\sqrt{h^{2}-\gamma^{2}}\right)  +\frac{1}{h}\sqrt{h^{2}-\gamma^{2}%
}\left[  1-\tanh^{2}\left(  \beta_{T}\sqrt{h^{2}-\gamma^{2}}\right)  \right]
\beta_{T}\right \}  \frac{\gamma}{\sqrt{h^{2}-\gamma^{2}}},\nonumber \\
\frac{\partial n_{y}}{\partial \gamma}  &  =0,\nonumber \\
\frac{\partial n_{z}}{\partial \gamma}  &  =\frac{1}{h},
\end{align}
for $h>\gamma$ and%
\begin{align}
\frac{\partial n_{x}}{\partial \gamma}  &  =0,\nonumber \\
\frac{\partial n_{y}}{\partial \gamma}  &  =-\frac{1}{\sqrt{1-(\frac{h}{\gamma
})^{2}}}\frac{h^{2}}{\gamma^{3}},\nonumber \\
\frac{\partial n_{z}}{\partial \gamma}  &  =-\frac{h}{\gamma^{2}},
\end{align}
for $h<\gamma$.

Fig.2 shows the result of $\frac{\partial \vec{n}}{\partial \gamma}$. We can see
that $\frac{\partial n_{x}}{\partial \gamma}$ depends on $\gamma$ and
temperature $T$, and $\frac{\partial n_{y}}{\partial \gamma}$ and
$\frac{\partial n_{z}}{\partial \gamma}$\ depends only on $\gamma$. Moreover,
at finite temperature $T\neq0$, $\frac{\partial \vec{n}}{\partial \gamma}$ is
discontinued at the $\gamma=h$. It means a "second-order" phase transition at
the EP: there exists a $\nu=\frac{1}{2}$ critical rule for $\frac{\partial
\vec{n}}{\partial \gamma}$:
\begin{equation}
\frac{\partial n_{x}}{\partial \gamma}=2\beta_{T}%
\end{equation}
\ for $h\rightarrow \gamma+0^{+}$ in the phase with $\mathcal{PT}$-symmetry
($h\geq \gamma$) and
\begin{equation}
\frac{\partial n_{y}}{\partial \gamma}=-(2h)^{^{-\frac{1}{2}}}(\gamma-h)^{-\nu}%
\end{equation}
\ for $\gamma \rightarrow h+0^{+}$ in the phase with $\mathcal{PT}$-symmetry
breaking ($h<\gamma$). While at zero temperature $T=0$,
\begin{equation}
\frac{\partial n_{x}}{\partial \gamma}=(2h)^{-\frac{1}{2}}(h-\gamma)^{-\nu}%
\end{equation}
\ for $h\rightarrow \gamma+0^{+}$ and
\begin{equation}
\frac{\partial n_{y}}{\partial \gamma}=-(2h)^{-\frac{1}{2}}(\gamma-h)^{-\nu}%
\end{equation}
\ for $\gamma \rightarrow h+0^{+}$. The results illustrate a \emph{zero
temperature anomaly} for the thermodynamic phase transition at EP!
\begin{figure}[ptb]
\includegraphics[clip,width=0.9\textwidth]{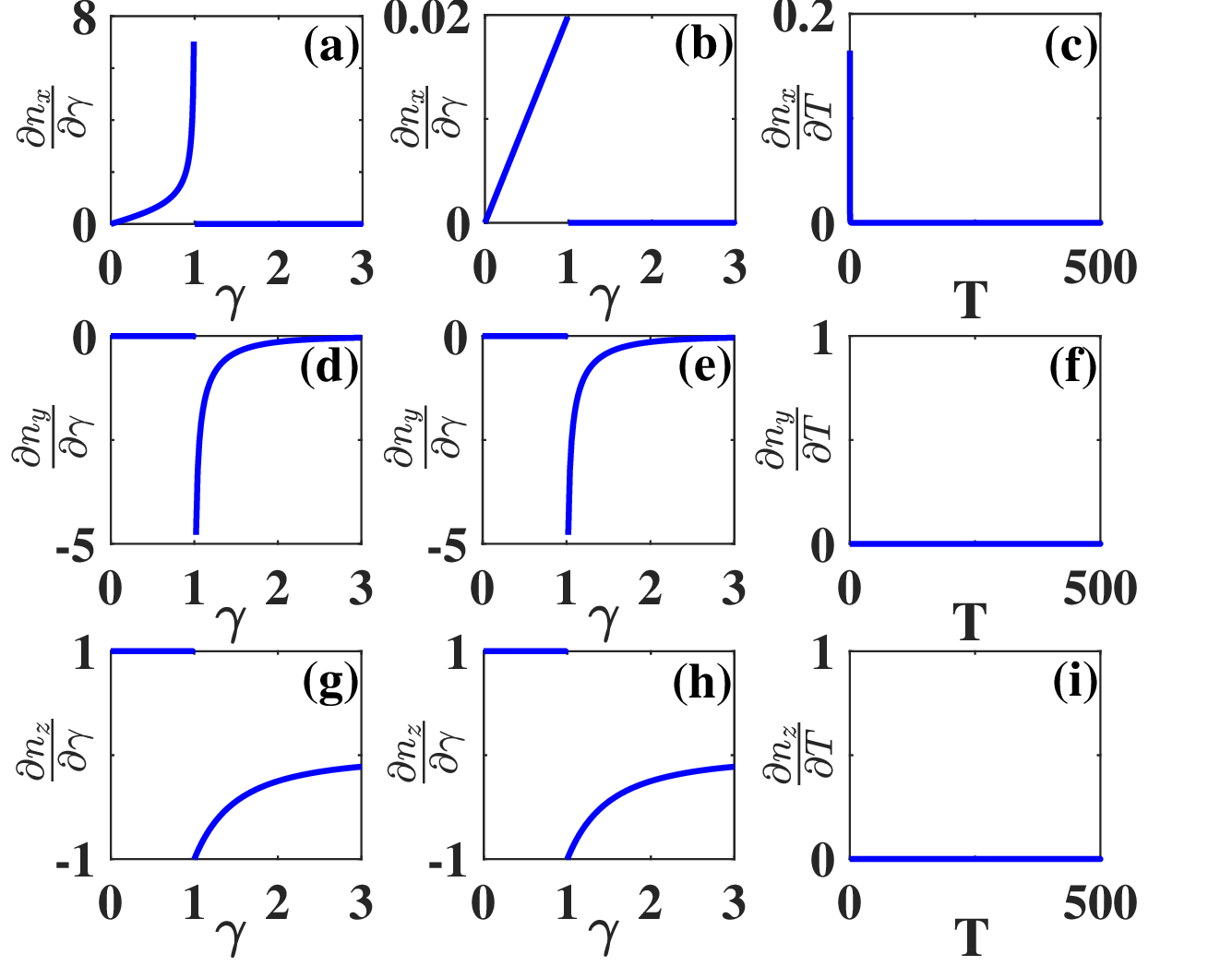}\caption{The derivatives
of spin average values $\frac{\partial \vec{n}}{\partial \gamma}=\frac
{\partial \left \langle \vec{\sigma}\right \rangle }{\partial \gamma}$ of the
thermal $\mathcal{PT}$ system versus $\gamma$ or temperature $T$. (a), (d) and
(g) are $\frac{\partial \left \langle \sigma_{i}\right \rangle }{\partial \gamma}$
versus $\gamma$ at the temperature $k_{B}T=0.01h$, respectively. (b), (e) and
(h) are $\frac{\partial \left \langle \sigma_{i}\right \rangle }{\partial \gamma}$
versus $\gamma$ at the temperature $k_{B}T=100h$, respectively. (c), (f) and
(i) are $\frac{\partial \left \langle \sigma_{i}\right \rangle }{\partial \gamma}$
versus temperature $T$ at $\gamma=0.999h$, respectively. }%
\end{figure}